\documentclass[11pt]{article}
\usepackage{graphics}
\usepackage{amsthm}
\usepackage{amsmath}
\usepackage{amssymb}
\usepackage{cite}

\newcommand{\st}{\rm such\ that}
\newcommand{\Eins}{\mbox{$1 \hspace{-1.0mm}  {\bf l}$}}
\newcommand{\Ket}[1]{|#1\rangle}
\newcommand{\Bra}[1]{\langle#1|}
\newcommand{\BraKet}[2]{\langle #1|#2\rangle}
\newcommand{\vc}[1]{\vec{ #1 }}
\newcommand{\e}{\varepsilon}
\newcommand{\0}{0}
\newcommand{\rank}[1]{\mbox{r}\lbrace #1 \rbrace}
\newcommand{\C}{{\cal C}}
\newcommand{\up}{\uparrow}
\newcommand{\down}{\downarrow}
\newcommand{\fc}[1]{f^{\dagger}_{#1}}
\newcommand{\bc}[1]{b^{\dagger}_{#1}}
\newcommand{\ba}[1]{b_{#1}}
\newcommand{\T}{{\rm T}}
\newcommand{\Tr}{{\rm Tr}}
\newcommand{\as}[1]{{\cal A}\left(#1\right)}
\newcommand{\hil}{{\cal H}}
\newcommand{\set}[1]{\left\{#1\right\}}
\newcommand{\must}{\ensuremath{\overset{!}{=}}}

\newtheorem{lemma}{Lemma}[section]
\newtheorem{Def}{Definition}[section] 
\newtheorem{Thm}{Theorem}[section] 
\newtheorem{Lem}{Lemma}[section]
\newtheorem{Prop}{Proposition}[section] 
\newtheorem{Crit}{Criterion}[section]
\newtheorem{Observ}{Observation}[section] 

\newtheorem*{bosdef1}{Definition~\ref{fermdef1}'}
\newtheorem*{bosthm1}{Theorem~\ref{fermthm1}'}
\newtheorem*{bosprop1}{Proposition~\ref{fermprop1}'}
\newtheorem*{bosdef2}{Definition~\ref{fermdef2}'}
\newtheorem*{boscrit1}{Criterion~\ref{fermcrit1}'}
\newtheorem*{bosdef3}{Definition~\ref{fermdef3}'}
\newtheorem*{boslem1}{Lemma~\ref{eqn:slaterwferm}'}
\newtheorem*{bosprop2}{Proposition~\ref{fermprop2}'}
\begin{document}

\title{Quantum Correlations in Systems of Indistinguishable Particles}

\author{K. Eckert$^1$, J. Schliemann$^{2,3}$, D. Bru\ss$^1$ and M.
Lewenstein$^1$\\\\
\it{$^1$
Institut f\"ur Theoretische  Physik,
Universit\"at  Hannover, 30167 Hannover, Germany}\\
\it{$^2$ Department of Physics, The University of Texas, Austin,
TX 78712}\\
\it{$^3$ Department of Physics and Astronomy, University of Basel, 4056
Basel, Switzerland}}

\maketitle

\begin{abstract}
We discuss quantum correlations in systems of indistinguishable particles in relation to
entanglement in composite quantum systems consisting of well separated subsystems.
Our studies are motivated by recent experiments and theoretical investigations on quantum
dots and neutral atoms in microtraps as tools for quantum information processing.
We present analogies between distinguishable particles, bosons and fermions in
low-dimensional Hilbert spaces. We introduce the notion of Slater rank for pure states of
pairs of fermions and bosons in analogy to the Schmidt rank for pairs of distinguishable 
particles. This concept is generalized to mixed states and provides a correlation
measure for indistinguishable particles.
Then we generalize these notions to pure fermionic and bosonic states in
higher-dimensional Hilbert spaces and also to the multi-particle case.
We review the results on quantum correlations in mixed fermionic states
and discuss the concept of fermionic Slater witnesses. Then the theory
of quantum correlations in mixed bosonic states and of bosonic Slater
witnesses is formulated. In both cases we provide methods of constructing optimal
Slater witnesses that detect the degree of quantum correlations in mixed fermionic
and bosonic states.
\end{abstract}

\tableofcontents



\section{Introduction}

The understanding and characterization of quantum entanglement is one 
of the most fundamental issues of modern quantum theory \cite{Per:95,BEZ:00},
and a lot of work has been devoted to this topic in the recent years
\cite{LEWrev:2000,H3rev:2001,TERrev:2001,BRUSS:2001}.

In the beginning of modern quantum theory, the notion of 
entanglement was first noted by Einstein, Podolsky, and Rosen \cite{EPR:35}, 
and by Schr\"odinger \cite{Sch:35}. While in those days quantum
entanglement and its predicted physical consequences were (at least partially)
considered as an unphysical property of the formalism (a ``paradox''),
the modern perspective on this issue is very different. Nowadays quantum
entanglement is seen as an experimentally verified property of nature,
that provides a resource for a vast variety of novel phenomena and concepts
such as quantum computation, quantum cryptography, or quantum teleportation.
Accordingly there are several motivations to study the
entanglement of quantum states:
\renewcommand{\labelenumi}{\Roman{enumi}.}
\begin{enumerate}
\item Interpretational and philosophical motivation: Entanglement plays an  
essential role in apparent ``paradoxes'' and counter-intuitive consequences 
of quantum mechanics \cite{EPR:35,Sch:35,BELL,WHE:1983}.
 
\item Fundamental physical motivation: 
The characterization of entanglement is one of the most fundamental 
open problems of quantum mechanics. It should answer the question what is the 
nature of quantum correlations in composite systems \cite{Per:95}.

\item Applied physical motivation: 
Entanglement plays an essential role in applications of quantum mechanics to 
quantum information processing, and in particular to quantum computing
\cite{qcrev}, 
quantum cryptography \cite{Eke:91,GISIN} and quantum communication
\cite{NIELSEN}(i.e.\ teleportation
\cite{BBC:1993,BBPS+:96} and super dense coding \cite{BeWi:96}). 
The resources needed to  
implement a particular protocol of quantum information processing  are 
closely linked to the entanglement properties of the states used  in the 
protocol. In particular, entanglement lies at the heart of quantum computing
\cite{BEZ:00}.

\item Fundamental mathematical motivation: 
The entanglement problem is directly related to one of the most challenging 
open problems of linear algebra and functional analysis: 
Characterization and classification of positive maps on ${\cal C}^*$ 
algebras \cite{LEWrev:2000,H3rev:2001,TERrev:2001,Jam:72,H3PPT:1996,Ter:00} (for mathematical literature see
\cite{STR:1963,WORO:1976,CHOI:1975,CHOI:1982}).
\end{enumerate}

While entanglement plays an essential role in quantum communication
between parties separated by macroscopic distances, the
characterization of quantum correlations at short distances is also an
open problem, which has received much less attention so far.
In this case the indistinguishable character of the particles
involved (electrons, photons,...) has to be taken into account. 
In his classic book, Peres \cite{Per:95} discussed the entanglement 
in elementary states of indistinguishable particles.
These are symmetrizations and antisymmetrizations of product states for bosons 
and fermions, respectively. It is easy to see that all such
states of two-fermion systems, and as well all such states 
formed by two non-collinear single-particle states 
in two-boson systems, are necessarily entangled in the usual sense.
However, in the case
of particles far apart from each other, this type of 
entanglement is not of physical relevance:   
{\sl ``No quantum prediction, referring to an atom located 
in our laboratory, is affected by the mere presence of similar atoms in remote
parts of the universe''}\cite{Per:95}. 
This kind of entanglement between indistinguishable
particles being far apart from each other 
is not the subject of this paper. Our aim here is rather to classify and 
characterize the quantum correlations between indistinguishable 
particles at short distances.
We discuss below why this problem is relevant for 
quantum information processing in various physical systems.
Perhaps the first attempt to study such quantum correlations in macroscopic
systems was done by A.J.\ Legett \cite{leggett:1980}. More recently he
has formulated the concept of {\it disconnectivity} \cite{leggett:2002}
of quantum states which is somewhat related to the concepts developed
in this review.

This paper is organized as follows: 
In section \ref{corrent} we illustrate the consequences of indistinguishability.
In section \ref{analogs} we describe
analogies between quantum correlations in systems of two indistinguishable
fermions, bosons, and two distinguishable parties where we concentrate on the
lowest-dimensional Hilbert spaces that allow for non-trivial correlation effects.
We derive the fermion and bosons analogues of recent results by Wootters
\cite{Woo:98}, by Kraus and Cirac \cite{KrCi:00}  
and by Khaneja and Glaser \cite{KhGl:00}. 
Our results shed new light on a 
question posed recently by Vollbrecht and Werner: {\sl Why two qubits are
special} \cite{VoWe:00} (see also \cite{AUD:2001}).
In section \ref{Kai} we report further
results on quantum correlations in pure states of indistinguishable fermions in 
higher-dimensional cases. Results on mixed fermionic states are summarized
in section \ref{sumferm}, and in section \ref{bosons} we report further 
results on identical bosons. We conclude in section \ref{concl}.



\section{Quantum correlations and entanglement}
\label{corrent}
\subsection{Physical systems: Quantum dots and neutral atoms in microtraps}

Semiconductor quantum dots \cite{SLM:00}~are a promising
approach to the physical realization of quantum computers. In these
devices charge carriers (e.g.~electrons) are confined in all three
spatial dimensions. Their electronic spectrum consists of discrete energy
levels since the confinement is of the order of the Fermi wavelength.
It is experimentally possible to control the number of electrons in
a such a dot starting from zero (e.g. in a GaAs heterostructure
\cite{tarucha:1996}).

When one wants to use quantum dots for quantum computation it is necessary
to define how the qubit (i.e. the basic unit of information)
should be physically realized. E.g.~the orbital electronic degrees of
freedom or the electron spin $\vc{S}$ can be chosen to form the qubit. An advantage
of the latter approach is that the decoherence time is much longer for the spin
than for the orbital degree of freedom (usually three orders of magnitude 
\cite{SLM:00,barenco:1995}).

The implementation of quantum algorithms needs single qubit and
two qubit quantum gates \cite{GRUSZ}. For the spin degree of
freedom the former
can be achieved by the application of a magnetic field exclusively to
a single spin \cite{DiLo:98}. It is well-known \cite{deutsch:1995} that
arbitrary computations can be done if, apart from single qubit rotations,
a mechanism by which two qubits can be entangled is available
(the entangling gate $\sqrt{{\rm SWAP}}$ \label{def:sqrtswap},
together with single qubit rotations, can be used to produce the fundamental
controlled-NOT gate). It was proposed in \cite{DiLo:98} to realize
this mechanism by temporarily coupling two spins in two dots.
The coupling, described by a Heisenberg Hamiltonian $H(t)=J(t)\vc{S}_1\vc{S}_2$,
can be turned on and off by lowering and raising the tunnel barrier between
neighboring quantum dots.

Another interesting type of physical implementation possibilities are neutral
atoms in magnetic \cite{mag} or optical \cite{opt1} microtraps.
Here each single neutral atoms is
trapped in a harmonic potential and their collisional interaction can be controlled by
temporarily decreasing the distance of the traps or by
state-selective switching of the trapping potential.

%
%
\subsection{Consequences of indistinguishability}\label{sec:exTwoQuantumDots}
We will use a schematic view of two electrons located in a double-well potential
to illustrate the consequences of indistinguishability for entanglement.
This description applies to the discussed examples of quantum information
processing in quantum dots and in optical or magnetical microtraps (replacing
electrons by atoms).
For this illustration we will assume that the qubit is modeled by the spin degree of freedom,
which we will denote by $\Ket{\up}$ and $\Ket{\down}$.  Furthermore we have two
spatial wavefunctions labeled $\Ket{\phi}$ and $\Ket{\chi}$,
initially localized in the left and in the right potential well, respectively.
Then the complete state-space is four dimensional:
$\{\Ket{\phi}\Ket{\up},\,\Ket{\chi}\Ket{\up},\,\Ket{\phi}\Ket{\down},\,\Ket{\chi}\Ket{\down}\}$.

We start with a situation where we have one electron in each well.
Even if they are prepared completely independently, their pure quantum state
has to be written in terms of Slater determinants in order to respect the indistinguishability.
Operator matrix elements between such single Slater
determinants contain terms due to the antisymmetrization of coordinates
(``exchange contributions'' in the language of Hartree-Fock theory).
However, if the moduli of $\BraKet{\vec{r}}{\phi}$, $\BraKet{\vec r}{\chi}$ have only
vanishingly small overlap, these exchange correlations will also tend
to zero for any physically meaningful operator. This situation is generically
realized if the supports of the single-particle wavefunctions are essentially
centered around locations being sufficiently apart from each other, or the
particles are separated by a sufficiently large energy barrier. 
In this case the antisymmetrization has no physical effect and for all practical
purposes it can be neglected.

Such observations clearly justify the treatment of indistinguishable particles
separated by macroscopic distances as effectively distinguishable objects.
So far, research in quantum information theory has concentrated on this case,
where the exchange statistics of particles forming quantum registers could be
neglected, or was not specified at all.

\begin{figure}[htbp]
\begin{center}
\includegraphics{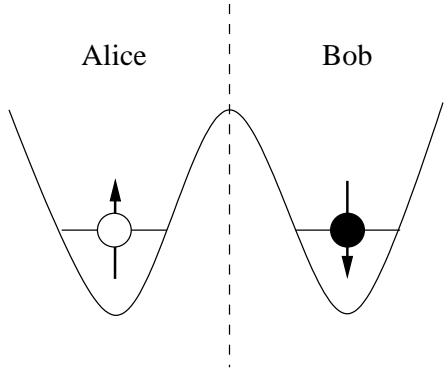} 
\end{center}
\caption[Initial state for the case of distinguishable particles]
{The initial state $\Ket{\psi_{\rm init}}_{AB}=\Ket{\phi}\Ket{\up}_A\otimes\Ket{\chi}\Ket{\down}_B$
i.e.~the case of separated wells. $\Ket{\phi}$ and
$\Ket{\chi}$ denote the spatial part of the wave-function
localized in the left and in the right well, respectively.}
\label{fig:2dotst0}
\end{figure}

Under these conditions we write an initial state
$\Ket{\psi_{\rm init}}_{AB}=\Ket{\phi}\Ket{\up}_A\otimes\Ket{\chi}\Ket{\down}_B$
where $A$ (Alice) and $B$ (Bob) are (physical meaningful) labels for the particle
in the left and the right dot, respectively.
The situation is shown in figure \ref{fig:2dotst0}.

\begin{figure}[hb]
\begin{center}
\includegraphics{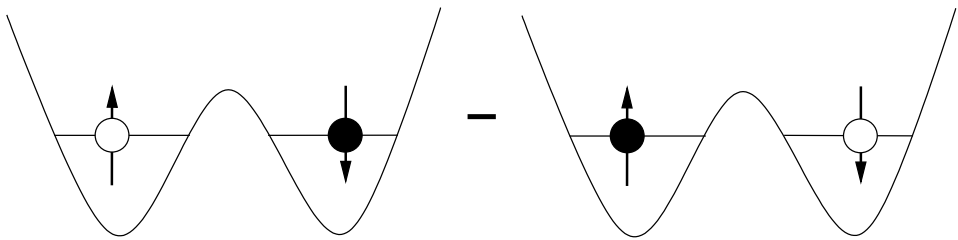} 
\end{center}
\caption[Illustration of the initial state with lowered tunnel barrier]
{Illustration of $\Ket{\psi(t_1)}$, i.e.~after lowering the tunnel barrier.
The electronic wavefunctions are no longer completely localized in one of the wells.}
\label{fig:2dotst1}
\end{figure}

Now we want to analyze the situation when the two wells have been moved closer together or
the energy barrier has been lowered. In such a situation the probability of finding, e.g.,
Alice's electron in the right well is non-vanishing.
Then the fermionic statistics is clearly essential and the
two-electron wave-function has to
be antisymmetrized and reads
$\Ket{\psi(t_1)} = \frac1{\sqrt{2}} \left[\,\,\Ket{\phi}\Ket{\up}_1\otimes\Ket{\chi}\Ket{\down}_2 - 
\Ket{\chi}\Ket{\down}_1\otimes\Ket{\phi}\Ket{\up}_2 \,\,\right]$.
The indices $A$ and $B$ are changed to $1$ and $2$ here to stress
that the enumeration of the particles is completely arbitrary since these
labels are not physical: because of the spatial overlap of the wavefunctions
the individual particles labeled '$1$' or '$2$' are not accessible independently.
The situation is shown in figure \ref{fig:2dotst1}.

Note that not only the labeling of the particles but also the notation suggesting
a tensor product structure of the space of states is misleading because the
actual state space is just a subspace of the complete tensor product \cite{Zan:01}. As a
consequence of this fact the antisymmetrized
state $\Ket{\psi(t_1)}$ formally resembles an entangled state although it is clear that this
entanglement is not accessible, and therefore cannot be used as a resource in the sense discussed
above for distinguishable particles.
To emphasize this fundamental difference between distinguishable and
indistinguishable particles, we will 
use the term \emph{quantum correlations} to characterize \emph{useful}
correlations in systems of indistinguishable particles as opposed to {\em correlations}
arising purely from their statistics (thereby following \cite{SLM:00}). 

Quantum correlations in systems of indistinguishable fermions
arise if more than one Slater determinant is involved, i.e.~if
there is no single-particle basis such that a
given state of $N$ indistinguishable fermions can be
represented  as an elementary  Slater determinant (i.e. a fully antisymmetric
combination of $N$ orthogonal single-particle states).
These correlations are  the analogue of quantum entanglement in 
separated systems and are essential for quantum information processing
in non-separated systems.

\begin{figure}[htbp]
\begin{center}
\includegraphics{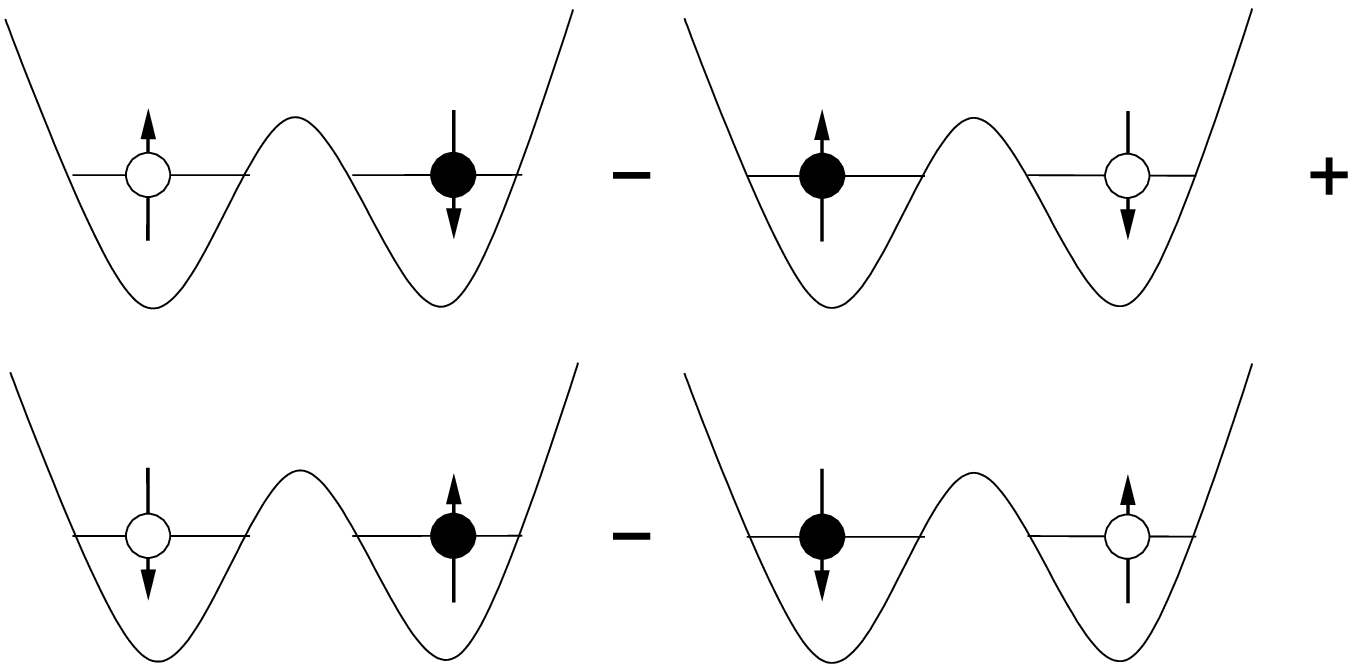}
\end{center}
\caption[Illustration of $\Ket{\psi(t_2)}$]{Illustration of $\Ket{\psi(t_2)}$}
\label{fig:2dotst2}
\end{figure}

As an example suppose it is possible to control the coupling $J(t)$ 
of the electrons such that at time $t_2$
$$
\Ket{\psi(t_2)}=\frac{1}{2} \left[\,
\Ket{\phi}\Ket{\up}_1\otimes\Ket{\chi}\Ket{\down}_2 - 
\Ket{\chi}\Ket{\down}_1\otimes\Ket{\phi}\Ket{\up}_2+
\Ket{\phi}\Ket{\down}_1\otimes\Ket{\chi}\Ket{\up}_2 - 
\Ket{\chi}\Ket{\up}_1\otimes\Ket{\phi}\Ket{\down}_2\, \right]
$$
which is illustrated in figure \ref{fig:2dotst2}.

\begin{figure}[htbp]
\begin{center}
\includegraphics{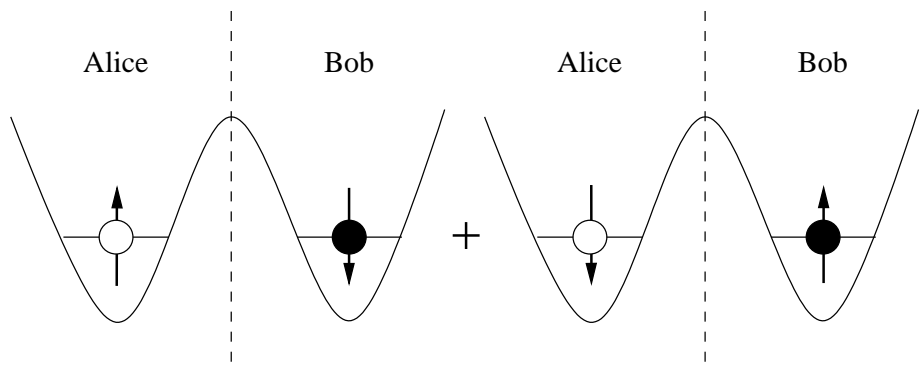} 
\end{center}
\caption[Illustration of the final state after separating the particles]
{The final state $\Ket{\psi_{\rm final}}_{AB}$. Raising the tunneling barrier 
localizes the wavefunctions again.}
\label{fig:2dotst3}
\end{figure}

In the given single-particle basis, $\Ket{\psi(t_2)}$ is written
in terms of two elementary Slater determinants (and evidently there is
no basis in which it can be written as a single one).
This state contains some useful correlations beyond the required permutation
symmetry as can be seen through localizing the particles again by switching off the
interaction, i.e.~raising the tunneling barrier or moving the wells apart
(here we neglect the effects of non-adiabaticity, see \cite{SLM:00} for a more
detailed study of these effects). This corresponds to a partition of the basis between
Alice and Bob, such that Alices Hilbert space is formed by
$\{\Ket{\phi}\Ket{\up},\,\Ket{\phi}\Ket{\down}\}$ and Bobs by  
$\{\Ket{\chi}\Ket{\up},\,\Ket{\chi}\Ket{\down}\}$.
Then again the electrons can be viewed as effectively distinguishable, provided that none
of the dots is occupied by two electrons. This does not happen here  because the final
final state is
$\Ket{\psi_{\rm final}}_{AB}=\frac1{\sqrt2} \left[\,\,\Ket{\phi}\Ket{\up}_A\otimes\Ket{\chi}\Ket{\down}_B+
\Ket{\phi}\Ket{\down}_A\otimes\Ket{\chi}\Ket{\up}_B\,\,\right]$, where
\emph{new} labels $A$ and $B$ are attributed to the particles, corresponding
to  the dot in which they are
found after separation, i.e. the electron found in the left (right) dot is named $A$
($B$). $\Ket{\psi_{\rm final}}_{AB}$ is shown in figure \ref{fig:2dotst3}.

The final state $\Ket{\psi_{\rm final}}_{AB}$,
shown in figure \ref{fig:2dotst3} is the Bell state $\Ket{\Psi^+}$,
i.e.~a \emph{maximally entangled}
two qubit state (thus the operation performed in this example is the entangling gate
$\sqrt{{\rm SWAP}}$).
In this sense it is reasonable to call $\Ket{\psi(t_2)}$ a
\emph{maximally correlated} state of two indistinguishable fermions in a four-dimensional
single-particle space and to view it as a resource for the production of entangled
states of distinguishable particles. 

Motivated by these considerations in \cite{SCK+:00} we have developed a
classification of states of two fermions with $2M$ accessible
single-particle states. This question was also addressed 
very recently by Li {\it et al.} \cite{LZLL:01} and by Paskauskas and You
\cite{PaYo:01}. In these papers two-boson systems are examined and
analogues to earlier results about two-fermion systems \cite{SLM:00,SCK+:00}
are derived. However, Refs. \cite{LZLL:01} and \cite{PaYo:01} differ 
in detail about which two-boson states should be considered
as analogues of entangled states (in a bipartite system) in a certain 
limiting case.

Zanardi \cite{Zan:01,Zan:02} discusses another approach, ignoring 
the original tensor product structure through a partition of the physical space into
subsystems. The entangled
entities then are no longer particles but modes. 
This approach may be seen as complementary to the
one followed here. For completeness we present the corresponding
formalism in appendix \ref{app:Zanardi}. It is reasonable to consider
both kinds of quantum correlations -- which one is more useful depends
on the particular situation, for instance on their usefulness for concrete
applications, e.g.\ cryptography or teleportation.



\section{Analogies between bosons, fermions, and distinguishable particles
in low-dimensional Hilbert spaces}
\label{analogs}

\subsection{Pure states: Schmidt rank and Slater rank}

\subsubsection{Schmidt rank of distinguishable particles}
The ``classic'' examples for quantum entanglement were studied in systems composed
of separated (and therefore distinguishable) subsystems. The most investigated
case involves two parties, say $A$(lice) and $B$(ob), having a 
finite-dimensional Hilbert space ${\cal H}_{A}$ and ${\cal H}_{B}$, 
respectively. This results in a total space 
${\cal H}={\cal H}_{A}\otimes{\cal H}_{B}$. An important tool for the
investigation of such bipartite systems is the bi-orthogonal Schmidt
decomposition \cite{Per:95}.
It states that for any state vector $|\psi\rangle\in{\cal H}$
there exist bases of ${\cal H}_{A}$ and ${\cal H}_{B}$ such that
\begin{equation}
|\psi\rangle=\sum_{i=1}^{r}z_{i}
\left(|a_{i}\rangle\otimes|b_{i}\rangle\right),\quad z_i>0\,\,\text{and}\,\,\sum_{i=1}^rz_i^2=1,
\label{Schmidt}
\end{equation}
where the basis states fulfill that
$\langle a_{i}|a_{j}\rangle=\langle b_{i}|b_{j}\rangle=\delta_{ij}$.
Thus, each vector in both bases for ${\cal H}_{A}$ and ${\cal H}_{B}$
occurs at most in only one product vector in the above expansion. 
The expression (\ref{Schmidt}) is an expansion of the state $|\psi\rangle$
into a basis of orthogonal product vectors with a minimum 
number $r$ of nonzero terms. This number can take values between one and
$\min\{\dim{\cal H}_{A},\dim{\cal H}_{B}\}$ and is called the
{\em Schmidt rank} of $|\psi\rangle$. $\Ket{\psi}$ is entangled if and only if $r>1$.

\subsubsection{Slater rank of fermionic states}
Let us now turn to the case of two identical fermions sharing an
$n$-dimensional single-particle space ${\cal H}_{n}$. The total  
Hilbert space is ${\cal A}({\cal H}_{n}\otimes{\cal H}_{n})$ where
${\cal A}$ denotes the antisymmetrization operator. A general state vector
can be written as
\begin{equation}
|w\rangle=\sum_{i,j=1}^{n}w_{ij}f^{\dagger}_{i}f^{\dagger}_{j}|0\rangle
\end{equation}
with fermionic creation operators $f^{\dagger}_{i}$ acting on the vacuum
$|0\rangle$. The antisymmetric coefficient
matrix $w_{ij}$ fulfills the normalization condition 
\begin{equation}
{\rm tr}\left( w^* w\right)=-\frac{1}{2}\,.
\end{equation}
Under a unitary transformation of the single-particle space,
\begin{equation}
f^{\dagger}_{i}\mapsto{\cal U}f^{\dagger}_{i}{\cal U}^{\dagger}=U_{ji}f^{\dagger}_{j}\,,
\end{equation}
$w$ transforms as 
\begin{equation}
w\mapsto UwU^{T}\,,
\end{equation}
where $U^{T}$ is the transpose of $U$. For any complex
{\em antisymmetric} $n\times n$ matrix $w$ there is a unitary transformation
$U$ such that $w'=UwU^{T}$ has nonzero entries only in $2\times 2$ 
blocks along the diagonal \cite{SCK+:00,Meh:77}. That is,
\begin{equation}
w'={\rm diag}\left[Z_{1},\dots,Z_{r},Z_{0}\right]
\quad{\rm with}\quad Z_i=\left[
\begin{array}{cc}
0 & z_i \\
-z_i & 0
\end{array}
\right]\,,
\label{z}
\end{equation}
where $z_{i}>0$ for $i\in\{1,\dots,r\}$, and $Z_{0}$ is the 
$(n-2r)\times(n-2r)$ null matrix.
Each $2\times 2$ block $Z_{i}$ corresponds to an elementary Slater 
determinant. Such elementary Slater determinants are the analogues
of product states in systems consisting of distinguishable parties.
Thus, when expressed in such a basis, the state $|w\rangle$ 
is a sum of elementary Slater determinants 
where each single-particle basis state occurs at most in one term.
This property is analogous to the bi-orthogonality of the 
Schmidt decomposition discussed above.
The matrix (\ref{z}) represents an expansion of $|w\rangle$ into a basis of
elementary Slater determinants with a minimum number $r$ of non-vanishing
terms. This number is analogous to the Schmidt rank for the distinguishable
case. Therefore we shall call it the {\em fermionic Slater rank} of 
$|w\rangle$ \cite{SCK+:00}, and an expansion of the above form a
{\em Slater decomposition} of $|w\rangle$. 

\subsubsection{Slater rank of bosonic states}
Similarly, a general state of a system of two indistinguishable bosons in
an $n$-dimensional single-particle space reads
\begin{equation}
|v\rangle=\sum_{i,j=1}^{n}v_{ij}b^{\dagger}_{i}b^{\dagger}_{j}|0\rangle
\end{equation}
with bosonic creation operators $b^{\dagger}_{i}$ acting on the vacuum state. The symmetric 
coefficient matrix $v_{ij}$ transforms under single-particle transformations 
just the same as in the fermionic case,
\begin{equation}
v\mapsto UvU^{T}\,.
\end{equation}
For any complex {\em symmetric} matrix $v$ there exists a unitary 
transformation
$U$ such that the resulting matrix $v'=UvU^{T}$ is diagonal 
\cite{LZLL:01,PaYo:01,Meh:77}, i.e.
\begin{equation}
v'={\rm diag}\left(z_{1},\dots,z_{r},0,\dots,0\right)
\label{bosSl}
\end{equation}
with $z_{i}\neq 0$ for $i\in\{1,\dots,r\}$. In such a single particle
basis the state $|v\rangle$ is a linear combination of elementary two-boson 
Slater permanents representing doubly occupied states. Moreover
Eq.~(\ref{bosSl}) defines an expansion of the given state $|v\rangle$ into
Slater permanents representing doubly occupied states
with the smallest possible number $r$ of
nonzero terms. We shall call this number the {\em bosonic Slater rank}
of $|v\rangle$. 

An expansion of the form (\ref{bosSl}) for a two-boson system
was also obtained very recently in Refs. \cite{LZLL:01,PaYo:01}. Moreover
the fermionic analogue (\ref{z}) of the bi-orthogonal
Schmidt decomposition of bipartite systems was also used earlier in
studies of electron correlations in Rydberg atoms \cite{atomphys}.

Regarding Slater determinant states in fermionic Hilbert spaces we also
mention interesting earlier work by Rombouts and Heyde 
\cite{Rombouts} who investigated
the question under what circumstances a given many-fermion wavefunction
can be cast as a Slater determinant built up from in
general {\em non-orthogonal} single-particle states. Since a wavefunction 
of this kind can in general not be written as a single Slater determinant
constructed from {\em orthogonal} single-particle states, i.e.\ has
non-trivial quantum correlations beyond simple antisymmetrization effects,
the criteria obtained in \cite{Rombouts} do not address the issues here.

As we saw, elementary Slater determinants in two-fermion systems, i.e.\
states with Slater rank one, are the 
natural analogues of product states in systems of distinguishable
parties.
One needs at least a fermionic state of slater rank two to form a quantum correlated
state corresponding to a Schmidt rank two state of separated particles. In contrast,
in the bosonic case one needs at least a state of Slater rank four to perform
the same task. This can already be seen from the fact that for distinguishable particles
entangled states need at least a $2\times2$ dimensional Hilbert space.

In the following we shall refer to the Schmidt rank or Slater rank of a
given pure state also as its {\em quantum correlation rank}.


\subsection{Magic bases, concurrence, and dualisation}
\label{sec:concurrence}

One of the most important issues in Quantum Information Theory is the
qualification and quantification of entanglement between several subsystems 
in a given state $\rho$.
For the case of two {\em distinguishable} parties, a useful measure of
entanglement of pure states is the von Neumann-entropy of reduced density 
matrices constructed from the density matrix $\rho=|\psi\rangle\langle\psi|$
\cite{BBP+:96} :
\begin{equation}
E(|\psi\rangle)=-{\rm tr}_{A}\left(\rho_{A}\log_{2}\rho_{A}\right)
=-{\rm tr}_{B}\left(\rho_{B}\log_{2}\rho_{B}\right)\,,
\label{vNeuEnt}
\end{equation}
where the reduced density matrices are obtained by tracing out one of the
subsystems, $\rho_{A}={\rm tr}_{B}\rho$ and vice versa. With the help of the 
bi-orthogonal Schmidt decomposition of $|\psi\rangle$ one shows that
both reduced density matrices have the same spectrum and therefore the 
same entropy, as stated in Eq.~(\ref{vNeuEnt}). In particular, the
Schmidt rank of $|\psi\rangle$ equals the algebraic rank of the reduced
density matrices. A pure state is non-entangled if and only if its 
reduced density matrices are again pure states,
and it is maximally entangled if its
reduced density matrices are ``maximally mixed'', i.e. if they have only
one non-zero eigenvalue with a multiplicity of
$\min\{\dim{\cal H}_{A},\dim{\cal H}_{B}\}$.

\subsubsection{Two distinguishable particles}
The lowest-dimensional system of two distinguishable particles having non-trivial
entanglement properties consists of just two qubits, i.e.\ 
$\dim{\cal H}_{A}=\dim{\cal H}_{B}=2$.
For this system, the entanglement measure (\ref{vNeuEnt}) takes a 
particularly simple form if the state $|\psi\rangle$ is expressed in the
so-called {\em magic basis},
\begin{eqnarray}
|\chi_{1}\rangle & = & \frac{1}{\sqrt{2}}
\left(|\uparrow\downarrow\rangle-|\downarrow\uparrow\rangle\right)\nonumber\\
|\chi_{2}\rangle & = & \frac{1}{\sqrt{2}}
\left(|\uparrow\uparrow\rangle+|\downarrow\downarrow\rangle\right)\nonumber\\
|\chi_{3}\rangle & = & \frac{i}{\sqrt{2}}
\left(|\uparrow\downarrow\rangle+|\downarrow\uparrow\rangle\right)\nonumber\\
|\chi_{4}\rangle & = & \frac{i}{\sqrt{2}}
\left(|\uparrow\uparrow\rangle-|\downarrow\downarrow\rangle\right)
\label{magbas}
\end{eqnarray}
i.e. $|\psi\rangle=\sum_{i=1}^{4}\alpha_{i}|\chi_{i}\rangle$,
where an obvious notation has been used. With respect to this
basis it holds \cite{HiWo:97,Woo:98}
\begin{equation}
E(|\psi\rangle)=h
\left(\frac{1+\sqrt{1-{\cal C}(|\psi\rangle)^{2}}}{2}\right)
\label{twoqubitent}
\end{equation}
where $h(x)=-x\log_{2}(x)-(1-x)\log_{2}(1-x)$ is the binary entropy
function and the ``concurrence'' ${\cal C}(|\psi\rangle)$ is defined by 
${\cal C}(|\psi\rangle)=|\sum_{i=1}^{4}\alpha_{i}^{2}|$.
Thus, a state is fully entangled if and only if all its coefficients
with respect to the magic basis have the same phase. What is furthermore
``magic'' about this basis is the fact that its elements are 
(pseudo-)eigenstates of the time reversal operator $\cal D$ \cite{Woo:98}:
\begin{equation}
{\cal D}|\chi_{i}\rangle=|\chi_{i}\rangle
\end{equation}
with 
\begin{equation}\label{twoqubitflip}
{\cal D}=((i\sigma^{y}_{A})\otimes(i\sigma^{y}_{B})){\cal K}\,.
\end{equation} 
Here $\sigma_{A,B}^{y}$ are Pauli matrices in the basis
used in the construction of the magic basis (\ref{magbas}),
and ${\cal K}$ is the operator of complex conjugation which acts on a product
state of basis vectors as ${\cal K}|\mu\nu\rangle=|\mu\nu\rangle$, where
$\mu,\nu\in\{\up,\down\}$, and on a general vector as
$K\sum_{\mu,\nu}\psi_{\mu\nu}\Ket{\mu\nu}=\sum_{\mu,\nu}\psi^*_{\mu\nu}\Ket{\mu\nu}$.
These relations are part of the definition of ${\cal K}$.
${\cal D}$ is invariant under arbitrary SU(2) transformations 
performed independently on the two 
subsystems, due to the relation $U\sigma^{y}U^{\T}=\sigma^{y}$ for any
$U\in{\rm SU(2)}$. It was pointed out in Ref.~\cite{VoWe:00} that this 
property is particular to the $2\times 2$ matrix $\sigma^{y}$, and does not
have a true analogue in higher dimensions. However, in the following we will
encounter similar invariance relations in higher-dimensional spaces
where the manifold of transformations
is restricted to a physically motivated subgroup of the unitary group.

Using the time reversal operator ${\cal D}$, the concurrence can be written 
as ${\cal C}(|\psi\rangle)=|\langle\tilde\psi|\psi\rangle|$ with 
$|\tilde\psi\rangle={\cal D}|\psi\rangle$. Moreover, since the entanglement
measure $E(|\psi\rangle)$ is a monotonous function of ${\cal C}(|\psi\rangle)$
with both functions ranging from zero to one, one  can equally well
use ${\cal C}(|\psi\rangle)$ as a measure of entanglement \cite{Woo:98,HHH:00},
as we shall do in the following.

The definition of the magic basis (and as well of the complex conjugation
operator $\cal K$) refers explicitly to certain bases in the two subsystems.
However, using the above invariance property, it is straightforward to show
that switching to different local bases has only trivial effects without
any physical significance. In particular, the concurrence is invariant under
such operations. This can be seen from writing $\Ket{\psi}$ in the computational
basis as $\Ket{\psi}=\sum_{\mu,\nu=\uparrow,\downarrow}\psi_{\mu,\nu}\Ket{\mu\nu}$. Then
${\cal C}(\Ket{\psi})=4 |\det(\psi)|^2$ and in this context it has been named ``tangle''
by Wootters \cite{Woo:98.2}.

\subsubsection{Two fermions}
Let us now turn to the case of two fermions. The lowest-dimensional system
allowing a Slater rank larger than one has a four-dimensional single-particle
space resulting in a six-dimensional two-particle Hilbert space.
This case was analyzed first in Ref.~\cite{SLM:00} where a
fermionic analogue of the two-qubit concurrence was found. It can be
constructed in the following way: For a given state 
$|w\rangle=\sum_{i,j=1}^{4}w_{ij}f^{\dagger}_{i}f^{\dagger}_{j}|0\rangle$
defined by its coefficient matrix $w_{ij}$, define the dual matrix
$\tilde w_{ij}$ by
\begin{equation}
\tilde w_{ij}=\frac{1}{2}\sum_{k,l=1}^{4}\varepsilon^{ijkl}w_{kl}^*\,,
\label{defdual}
\end{equation}
with $\epsilon^{ijkl}$ being the usual totally antisymmetric unit tensor.
Then the concurrence ${\cal C}(|w\rangle)$ can be defined as \cite{note1}
\begin{equation}
{\cal C}(|w\rangle)=|\langle\tilde w|w\rangle|=\left|\sum_{i,j,k,l=1}^{4}
\varepsilon^{ijkl}w_{ij}w_{kl}\right|
=|8 \left(w_{12}w_{34}+w_{13}w_{42}+w_{14}w_{23}\right)|\,.
\end{equation}
Obviously, ${\cal C}(|w\rangle)$ ranges from zero to one. 
Importantly it vanishes
if and only if the state $|w\rangle)$ has the fermionic Slater rank one, i.e.
is an elementary Slater determinant. This statement was proved first in
Ref.~\cite{SLM:00}; an alternative proof can be given using the 
Slater decomposition of $|w\rangle$ and observing that
\begin{equation}
\det w=\left(\frac{1}{8}\langle\tilde w|w\rangle\right)^{2}\,.
\label{det}
\end{equation}
This relation is just a special case of a general expression for
the determinant of an antisymmetric $(2K)\times(2K)$ matrix $w$,
\begin{equation}\label{eqn:asdet}
\det w=\Big(\frac{1}{2^{K}K!}\sum_{i_{1},\dots,i_{2K}=1}^{2K}
\varepsilon^{i_{1}\dots i_{2K}}w_{i_{1}i_{2}}\dots w_{i_{2K-1}i_{2K}}
\Big)^{2}\,,
\end{equation}
which can be proved readily by the same means. Note the formal analogy to
the definition of the tangle of two qubits \cite{Woo:98.2}. 

Moreover, it is straightforward to show that the concurrence is invariant
under arbitrary SU(4) transformations in the single-particle space.
More generally, for any two states $|w_{1}\rangle$, $|w_{2}\rangle$
being subject to a single-particle transformation ${\cal U}$,
$|w_{i}\rangle\mapsto|u_{i}\rangle={\cal U}|w_{i}\rangle$,
$i\in\{1,2\}$, it holds
\begin{equation}
\langle\tilde u_{2}|u_{1}\rangle=\det {\cal U}\langle\tilde w_{2}|w_{1}\rangle\,.
\end{equation}

In summary, the concurrence ${\cal C}(|w\rangle)$ constitutes,
in analogy to the case of two qubits, a 
measure of ``entanglement'' for any two-fermion state $|w\rangle$. 
It ranges from zero for ``non-entangled'' states (having Slater rank one)
to one for ``fully entangled'' states which are collinear with their duals.
In this case the Slater decomposition of $|w\rangle$ consists of two
elementary Slater determinants having the same weight.

From the occurrence of the complex conjugation in the definition of the
dual state (\ref{defdual}) it is obvious that the dualisation of a two-fermion 
state is an anti-linear operation. In fact, as we will illustrate below, 
the dualisation of a state $|w\rangle$ again corresponds to time reversal. As 
another physical interpretation, it can be identified with a particle-hole-transformation,
\begin{equation}
{\cal U}_{ph}f^{\dagger}_{i}{\cal U}^{\dagger}_{ph}=f_{i}\quad,\quad
{\cal U}_{ph}|0\rangle
=f^{\dagger}_{1}f^{\dagger}_{2}f^{\dagger}_{3}f^{\dagger}_{4}|0\rangle\,,
\label{ph1}
\end{equation}
along with a complex conjugation. The operator of dualisation ${\cal D}$, 
$|\tilde w\rangle={\cal D}|w\rangle$, can be written as
\begin{equation}
{\cal D}=-{\cal U}_{ph}{\cal K}
\label{defdualop}
\end{equation}
where $\cal K$ is the anti-linear operator of complex conjugation
acting on the single-particle states and the fermionic vacuum as
\begin{equation}
{\cal K}f^{\dagger}_{i}{\cal K}=f^{\dagger}_{i}\quad,\quad
{\cal K}f_{i}{\cal K}=f_{i}\quad,\quad{\cal K}|0\rangle=|0\rangle\,.
\label{defK}
\end{equation}

As a further analogy to the two-qubit case one also has a magic basis,
i.e. a basis of (pseudo-)eigenstates of the dualisation operator
(\ref{defdualop}):
\begin{eqnarray}
|\chi_{1}\rangle & = & \frac{1}{\sqrt{2}}
\left(f^{\dagger}_{1}f^{\dagger}_{2}+f^{\dagger}_{3}f^{\dagger}_{4}\right)|0\rangle\nonumber\\
|\chi_{2}\rangle & = & \frac{1}{\sqrt{2}}
\left(f^{\dagger}_{1}f^{\dagger}_{3}+f^{\dagger}_{4}f^{\dagger}_{2}\right)|0\rangle\nonumber\\
|\chi_{3}\rangle & = & \frac{1}{\sqrt{2}}
\left(f^{\dagger}_{1}f^{\dagger}_{4}+f^{\dagger}_{2}f^{\dagger}_{3}\right)|0\rangle\nonumber\\
|\chi_{4}\rangle & = & \frac{i}{\sqrt{2}}
\left(f^{\dagger}_{1}f^{\dagger}_{2}-f^{\dagger}_{3}f^{\dagger}_{4}\right)|0\rangle\nonumber\\
|\chi_{5}\rangle & = & \frac{i}{\sqrt{2}}
\left(f^{\dagger}_{1}f^{\dagger}_{3}-f^{\dagger}_{4}f^{\dagger}_{2}\right)|0\rangle\nonumber\\
|\chi_{6}\rangle & = & \frac{i}{\sqrt{2}}
\left(f^{\dagger}_{1}f^{\dagger}_{4}-f^{\dagger}_{2}f^{\dagger}_{3}\right)|0\rangle
\end{eqnarray}
Expressed in this basis, the concurrence of a state
$|w\rangle=\sum_{i=1}^{6}\alpha_{i}|\chi_{i}\rangle$ just reads
${\cal C}(|w\rangle)=|\sum_{i=1}^{6}\alpha_{i}^{2}|$. Therefore, $|w\rangle$ has
a concurrence of modulus one if and only if all its coefficients with respect
to the magic basis have the same phase.

Like in the two-qubit case, the definition of the complex conjugation operator
$\cal K$ and the magic basis refers to a certain choice of basis in the
single-particle space. However, due to the invariance properties described 
above, this does not have physically significant consequences.

A four-dimensional single-particle space can formally be viewed as the
Hilbert space of a spin-$\frac{3}{2}$-object. Having two indistinguishable 
fermions in this system means coupling two spin-$\frac{3}{2}$ states
to total spin states that are antisymmetric under particle exchange. This
leads to the two multiplets with even total spin, i.e. $S=2$ and $S=0$.
When the single-particle states labeled so far by $i\in\{1,2,3,4\}$ are 
interpreted as states of a spin-$\frac{3}{2}$ particle with $S^{z}=\frac{3}{2}-(i-1)$
($\hbar=1$) these multiplet states read explicitly:
\begin{eqnarray}
|2,2\rangle & = & f^{\dagger}_{1}f^{\dagger}_{2}|0\rangle\nonumber\\
|2,1\rangle & = & f^{\dagger}_{1}f^{\dagger}_{3}|0\rangle\nonumber\\
|2,0\rangle & = & 
\frac{1}{\sqrt{2}}\left(f^{\dagger}_{1}f^{\dagger}_{4}+f^{\dagger}_{2}f^{\dagger}_{3}\right)
|0\rangle\nonumber\\
|2,-1\rangle & = & f^{\dagger}_{2}f^{\dagger}_{4}|0\rangle\nonumber\\
|2,-2\rangle & = & f^{\dagger}_{3}f^{\dagger}_{4}|0\rangle\nonumber\\
|0,0\rangle & = & 
\frac{i}{\sqrt{2}}\left(f^{\dagger}_{1}f^{\dagger}_{4}-f^{\dagger}_{2}f^{\dagger}_{3}\right)|0\rangle
\end{eqnarray}
The phase of the last singlet state has been adjusted such that
the dualisation operator reads in this ordered basis
\begin{equation}
{\cal D}=\left(
\begin{array}{cccccc}
0 & 0 & 0 & 0 & 1 & 0 \\
0 & 0 & 0 & -1 & 0 & 0 \\
0 & 0 & 1 & 0 & 0 & 0 \\
0 & -1 & 0 & 0 & 0 & 0 \\
1 & 0 & 0 & 0 & 0 & 0 \\
0 & 0 & 0 & 0 & 0 & 1 
\end{array}
\right){\cal K}
\end{equation}
which is nothing but the time reversal operator acting on the two 
multiplets $S=2$ and $S=0$. Thus, the operation of dualisation can
also be interpreted as a time reversal operation performed on appropriate spin 
objects. However, this interpretation is not necessarily the physically most
natural one: the notion of dualisation and entanglement-like
quantum correlations between indistinguishable fermions was first
investigated for the case of two electrons (carrying a spin of $\frac{1}{2}$)
in a system of two quantum dots \cite{SLM:00}.
There the interpretation of $\cal D$ as
the anti-linear implementation of a particle-hole-transformation, rather than
time reversal of formal spin objects, seems clearly more appropriate.
Therefore we shall retain the general term dualisation for such an
operation in the two kinds of systems investigated so far, and also
in the case of a bosonic system to be explored below.

\subsubsection{Two bosons}
Having established all these analogies between two qubits and a system of
two fermions sharing a four-dimensional single particle space,
it is natural to ask whether similar observations can be made for a system
of two bosons. The bosonic system showing properties analogous to those
discussed before consists of two indistinguishable bosons in a 
two-dimensional single-particle space. This is the smallest-dimensional 
system admitting states with a bosonic Slater rank greater than one.
It can be viewed as the symmetrized version of the two-qubit system, and its
two-boson space represents therefore the Hilbert space of a spin-1-object.

A general state vector of this system reads
$|v\rangle=\sum_{i,j=1}^{4}v_{ij}b^{\dagger}_{i}b^{\dagger}_{j}|0\rangle$ with a
coefficient matrix
\begin{equation}
v=\left(
\begin{array}{cc}
A & B \\
B & C
\end{array}
\right)
\end{equation}
being subject to the normalization condition $2|A|^{2}+4|B|^{2}+2|C|^{2}=1$.
The appropriate dualisation operator is
\begin{equation}
{\cal D}={\cal R}{\cal K}
\end{equation}
where $\cal K$ is the complex conjugation operator with analogous properties
as above, and the operator $\cal R$ acts in the single-particle space as
\begin{equation}
b^{\dagger}_{a}\mapsto{\cal R}b^{\dagger}_{a}{\cal R}^{\dagger}
=\sum_{b=1}^{2}-i\sigma_{ba}^{y}b^{\dagger}_{b}\,,
\end{equation}
i.e. $\cal R$ exchanges the state labels inferring a sign. When expressed
in the ordered basis $(\,(1/\sqrt{2})b^{\dagger}_{1}b^{\dagger}_{1}|0\rangle,
b^{\dagger}_{1}b^{\dagger}_{2}|0\rangle,(1/\sqrt{2})b^{\dagger}_{2}b^{\dagger}_{2}|0\rangle\,)$
the dualisation operator reads
\begin{equation}
{\cal D}=\left(
\begin{array}{ccc}
0 & 0 & 1\\
0 & -1 & 0\\
1 & 0 & 0
\end{array}
\right){\cal K}
\end{equation}
which is, as natural, just the time reversal operator for a spin-1-object.
Now the concurrence ${\cal C}(|v\rangle)$ can be defined as the 
modulus of the scalar product
of $|v\rangle$ with its dual $|\tilde v\rangle={\cal D}|v\rangle$
and expressed as
\begin{equation}\label{eqn:conbos}
{\cal C}(|v\rangle)=|\langle\tilde v|v\rangle|=
\left|4\left(AC-B^{2}\right)\right|=|4\det v|\,.
\end{equation}
The last equation makes immediately obvious that ${\cal C}(|v\rangle)$ is
invariant under SU(2) transformations in the single-particle space and
is zero if and only if $|v\rangle$ has bosonic Slater rank one.
Therefore this quantity has properties that are completely analogous to
the two other cases before. A magic basis of the two-boson
space is given by
\begin{eqnarray}
|\chi_{1}\rangle & = & \frac{1}{2}
\left(b^{\dagger}_{1}b^{\dagger}_{1}+b^{\dagger}_{2}b^{\dagger}_{2}\right)|0\rangle\nonumber\\
|\chi_{2}\rangle & = & \frac{i}{2}
\left(b^{\dagger}_{1}b^{\dagger}_{1}-b^{\dagger}_{2}b^{\dagger}_{2}\right)|0\rangle\nonumber\\
|\chi_{3}\rangle & = & i\left(b^{\dagger}_{1}b^{\dagger}_{2}\right)|0\rangle\,.
\end{eqnarray}
Again, the concurrence is maximal, i.e.\ ${\cal C}(|v\rangle)=1$, if and only if
the coefficients of $|v\rangle$ in the magic basis have all the same
phase.


\subsection{Mixed states and unified correlation measure:
Wootters' formula and its analogues}

We now turn to the case of mixed states characterized by a density matrix
$\rho$.
For a general bipartite system (not necessarily consisting of just two
qubits) the {\em entanglement of formation} of a state $\rho$ is defined
by the expression \cite{BDS+:96,VP+:97}:
\begin{equation}
{\cal E}_{F}(\rho)=
{\rm inf}_{\{p_{i},|\psi_{i}\rangle\}}
\left\{\sum_{i}p_{i}E(|\psi_{i}\rangle)\right\}
\label{inf}
\end{equation}
where $E$ is the entanglement measure (\ref{vNeuEnt}) for pure states,
and the infimum is taken over all decompositions 
$\rho=\sum_{i}p_{i}|\psi_{i}\rangle\langle\psi_{i}|$ of the density matrix
$\rho$ in terms of normalized but in general not orthogonal states
$|\psi_{i}\rangle$ and positive coefficients $p_{i}$ with  $\sum_{i}p_{i}=1$.

In the case of two qubits an equivalent entanglement measure is,
following Wootters \cite{Woo:98}, given by
\begin{equation}
{\cal C}(\rho)={\rm inf}_{\{p_{i},|\phi_{i}\rangle\}}
\left\{\sum_{i}p_{i}{\cal C}(|\phi_{i}\rangle)\right\}.
\label{defcm}
\end{equation}
where the concurrence ${\cal C}$ enters directly instead of via the entropy 
expression (\ref{twoqubitent}). This possibility relies on the fact that 
$E(|\psi\rangle)$ for a pure state $|\psi\rangle$ is a monotonous 
function of the modulus of its concurrence, and the infimum in (\ref{inf}) is 
actually realized by a decomposition of $\rho$ where each state $|\psi\rangle$
has the same concurrence, such that the summation becomes trivial
\cite{Woo:98}.

Furthermore ${\cal C}$ can be given by a closed expression avoiding
the minimization entering (\ref{defcm}), namely
\begin{equation}
{\cal C}(\rho)=\max\{0,\lambda_{1}-\lambda_{2}-\lambda_{3}-\lambda_{4}\}\,,
\label{woo1}
\end{equation}
where $\lambda_{1},\lambda_{2},\lambda_{3},\lambda_{4}$ are, in descending
order of magnitude, the square roots of the singular values of the matrix 
$\rho\tilde\rho$ with $\tilde\rho={\cal D}\rho{\cal D}^{-1}$. 
The singular 
values of the (in general non-hermitian) matrix  $\rho\tilde\rho$ can 
readily be shown to be always real and non-negative.

The formula (\ref{woo1}) was first conjectured, and proved for a subclass
of density matrices, by Hill and Wootters \cite{HiWo:97}, 
and proved for the general case
shortly afterwards by Wootters \cite{Woo:98}.

The expression (\ref{defcm}) can obviously be read as a unified correlation 
measure for all of the three cases discussed above, namely two qubits,
two indistinguishable fermions in a four-dimensional single-particle space,
and finally two indistinguishable bosons in a two-dimensional single-particle space.
This unified measure ranges from
zero to one and vanishes for pure states with correlation rank one.
Moreover this quantity is invariant under ``local unitary transformations'', i.e. 
independent SU(2) transformation in the two qubit spaces, or, for
indistinguishable particles, unitary transformations of the
single-particle Hilbert space. Moreover, in all three systems, the
dualisation operation has the property that for any two pure states
$|\psi\rangle$, $|\chi\rangle$ and there duals $|\tilde\psi\rangle$,
$|\tilde\chi\rangle$, the identity
$\langle\tilde\psi|\chi\rangle=\langle\tilde\chi|\psi\rangle$ holds.

These three properties are sufficient to take Wootters' proof \cite{Woo:98}
over to the other two cases. Therefore, the correlation measure
${\cal C}(\rho)$ of a general mixed state of two fermions sharing a
four-dimensional single-particle space can be expressed as
\begin{equation}
{\cal C_F}(\rho)=\max\{0,\lambda_{1}
-\lambda_{2}-\lambda_{3}-\lambda_{4}-\lambda_{5}-\lambda_{6}\}\,,
\label{woo2}
\end{equation}
and the analogous expression for the two-boson system reads
\begin{equation}
{\cal C_B}(\rho)=\max\{0,\lambda_{1}-\lambda_{2}-\lambda_{3}\}\,.
\label{woo3}
\end{equation}
Again, in full analogy to the two-qubit case, the $\lambda_{i}$ are, in
descending oder of magnitude, the square roots of the 
singular values of of $\rho\tilde\rho$ with 
$\tilde\rho={\cal D}\rho{\cal D}^{-1}$ being the appropriately dualised 
density matrix. As in Ref.~\cite{Woo:98} one can show that the
singular values of $\rho\tilde\rho$ are always real and non-negative.

In Ref.~\cite{SCK+:00} we have given a somewhat alternative proof
of the Eqs.~(\ref{woo1}),  (\ref{woo2}),
(\ref{woo3}) in Ref.~\cite{SCK+:00} by discussing the matrix 
\begin{equation}
C_{ij}=\langle\tilde\psi_{i}|\psi_{j}\rangle
\end{equation}
which arises from the decompositions 
$\rho=\sum_{i}|\psi_{i}\rangle\langle\psi_{i}|$ of a given density
matrix in terms of (in general non-normalized) states $|\psi_{i}\rangle$.


\subsection{Invariance group of the dualisation and general
unitary transformations}

To simplify the notation we shall in the following denote analogous quantities
referring to the three systems above, namely two qubits, two fermions, and two 
bosons, in the form $[\cdot,\cdot,\cdot]$. As an example, the unified 
correlation measure reads 
\begin{equation}
{\cal C}(\rho)=\max\{0,\lambda_{1}-\sum_{i=2}^{d}\lambda_{i}\}\,,
\label{uniwoo}
\end{equation}
where $d=[4,6,3]$ is the dimension of the full Hilbert space of each system.

Let us denote the group of (special) unitary transformations of the
appropriate total Hilbert space by
$G=[{\rm SU(4)},{\rm SU(6)},{\rm SU(3)}]$, and the group
of ``local transformations'' (acting on the qubit spaces, or on the
appropriate single-particle space) by 
$H=[{\rm SU(2)}\otimes{\rm SU(2)},{\rm SU(4)},{\rm SU(2)}]$. 
In the following we do not
distinguish between $H$ and its $[4,6,3]$-dimensional representation in the
full Hilbert space.

For a given state $|\Psi\rangle$ expressed in the magic basis,
$|\Psi\rangle=\sum_{i=1}^{d}\alpha_{i}|\chi_{i}\rangle$, the action of the
dualisation operator is
\begin{equation}
{\cal D}|\Psi\rangle=\sum_{i=1}^{d}\alpha^*_{i}|\chi_{i}\rangle\,.
\end{equation}
Therefore, within the magic basis, the dualisation operator ${\cal D}$
acting on the coordinate vector with components $\alpha_{i}$
is just ${\cal D}={\bf 1}_{d\times d}{\cal K}$.
We now investigate the subgroup $M$ of $G$ consisting of all 
unitary transformations 
$\cal U$ that leave scalar products of the form 
$\langle\tilde\Psi_{1}|\Psi_{2}\rangle$ invariant, i.e.
$\langle\tilde\Phi_{1}|\Phi_{2}\rangle=\langle\tilde\Psi_{1}|\Psi_{2}\rangle$
for $|\Psi_{i}\rangle\mapsto|\Phi_{i}\rangle={\cal U}|\Psi_{i}\rangle$,
$i\in\{1,2\}$ with two  arbitrary states $|\Psi_{i}\rangle$. Such
transformations must leave the dualisation operator invariant,
\begin{equation}
{\cal U}{\cal D}{\cal U}^{\dagger}={\cal D}
\label{orthocond}
\end{equation}
which means in a magic basis
\begin{equation}
{\cal U}{\cal U}^{T}={\bf 1}\,.
\end{equation}
It follows that ${\cal U}$ is, in a magic basis, real and orthogonal, 
${\cal U}\in{\rm O}(d)$, $d=[4,6,3]$, and,
since $\det{\cal U}=1$, ${\cal U}\in{\rm SO}(d)$.
In a general basis ${\cal U}$ is an element of the
corresponding equivalent representation of ${\rm SO}(d)$.
Thus, $M$ is just the special orthogonal group ${\rm SO}(d)$. 
Moreover, since $H$ is certainly contained in $M$, and
${\rm SO}(d)$ is in all three cases a representation of $H$, we have
$M=H$. This result can of course also be achieved by a straightforward
construction of $M$ in terms of its generators.

The fact that the group of ``local transformations'' is in all three
cases identical to the unitary invariance group of the dualisation operator
is peculiar to these small system studied here. As an example, consider
a system of $2n$ qubits. The generalization of the operator
$\cal D$ is obvious, and, since the number of qubits is even, one also finds
magic bases. Then the argument leading to the condition (\ref{orthocond})
can be repeated in a straightforward way, with the conclusion that the
invariance group of $\cal D$ is essentially ${\rm SO}(2^{2n})$, containing
$2^{2n-1}(2^{2n}-1)$ independent parameters. In contrast, $H$ consists
just of copies of SU(2) and has $6n$ parameters. Thus only for $n=1$
we find $M=H$; for larger $n$ $H$ is a proper subgroup of $M$.
Similar arguments can be made for the generalization of the
other two-particle systems to larger Hilbert spaces.

As we saw, $H$ contains in all of the three cases above
all unitary transformations which turn the states of  
a given magic basis into another magic basis.
This result is the key ingredient
to prove a recent result by Kraus and Cirac for the two-qubit case
\cite{KrCi:00} (also derived independently Khaneja and Glaser 
\cite{KhGl:00}).
Here we shall just state the general result: Any unitary
transformation ${\cal U}\in G$ of the full Hilbert space can be written 
in any magic basis as
\begin{equation}
{\cal U}={\cal V}_{1}{\cal U}_{d}{\cal V}_{2}
\label{transdecomp}
\end{equation}
with ${\cal V}_{1},{\cal V}_{2}\in H$ and ${\cal U}_{d}\in G$ diagonal.
Thus, ${\cal U}_{d}$ contains $d-1=[3,5,2]$ 
independent parameters in the phase
factors on its diagonal. Since the operations ${\cal V}_{1},{\cal V}_{2}$
do not change the concurrence of any state, Eq.~(\ref{transdecomp})
constitutes a decomposition of a unitary transformation into parts
which preserve ``entanglement'' and a part that does not.
Therefore this decomposition is very helpful in the study of
generation and deletion of entanglement in the course of a unitary
transformation \cite{KrCi:00}. 

The proof of the statement made in Eq.~(\ref{transdecomp}) for all
three cases is completely analogous to the one given in the appendix
of Ref.~\cite{KrCi:00} for the two-qubit case. For all systems
the dimensions of the groups involved on the r.h.s. of
Eq.~(\ref{transdecomp}) add up to give the dimension of the l.h.s.
For analogous systems with larger Hilbert spaces such a decomposition of
unitary transformations is in general not possible, as one can see from
similar dimensional arguments as given above.


\subsection{Summary}

In this section we have studied three different two-body systems:
\begin{itemize}
\item Two distinguishable qubits with a four-dimensional total Hilbert space.
\item Two indistinguishable fermions with a four-dimensional
single-particle Hilbert space. The total two-particle space is six-dimensional.
\item Two indistinguishable bosons in a two-dimensional single-particle
space. The total two-particle space is three-dimensional.
\end{itemize}
These three systems share a series of properties:
\begin{itemize}
\item They represent the smallest systems (regarding
the dimensions of the Hilbert spaces) which allow for nontrivial correlation
effects, i.e. a correlation rank larger than one.
\item In all three cases there is a {\em correlation measure}, the concurrence,
related to an appropriately defined time reversal operator (dualisation).
In the fermionic case this operator can be also interpreted as the
anti-linear implementation of a particle-hole-transformation.
\item In all three cases a {\em magic basis} exists that consists of states
being invariant under the dualisation operator.
Maximally correlated states (with a concurrence of one) 
have real coefficients (up to a common phase factor) when expressed in
the magic basis.
\item In the fermionic and bosonic case there are perfect analogues of 
Wootters' formula \cite{Woo:98} for the correlation
measure of mixed two-qubit states. 
\item In all three systems a general unitary transformation ${\cal U}$ of
the full Hilbert space can be written (in the magic basis) as
\begin{equation}
{\cal U}={\cal V}_{1}{\cal U}_{d}{\cal V}_{2}
\end{equation}
where ${\cal U}_{d}$ is diagonal and ${\cal V}_{1},{\cal V}_{2}$
are appropriate single-particle transformations. The result
for the two-qubit case was derived recently in Refs. 
\cite{KrCi:00,KhGl:00}.
\item In all three cases the group of single-particle transformations and
the group of unitary invariance transformations of
the dualisation operator coincide. In 
higher-dimensional cases the invariance group of the dualisation operator
is larger than the group of single-particle transformations.
\end{itemize}
Recently Verstraete {\it et al.} derived an explicit expression for a global
unitary transformation which maximizes the entanglement of a given mixed state
of two qubits\cite{VAD:01}. Given the analogies derived above, we conjecture that
the results of ref.\ \cite{VAD:01} can be applied to the other two systems studied here.

In fact all these analogies can be derived from the following observation: as explained
a two-fermion system in a four-dimensional single-particle space can be realized by two 
electrons on two neighboring quantum dots, taking into account only the lowest
orbital states and the spin degree of freedom. After a projection onto the subspace where
each dot contains one and only one electron a two-qubit system remains. Finally we arrive
at a system of two bosons in a two-dimensional single-particle space by symmetrization.
In this sense the system of two fermions is the most general one. Therefore,
 the question ``{\sl Why two qubits are special}'', posed recently by Vollbrecht and
Werner\cite{VoWe:00},
can be formulated more generally: {\sl What is special about two fermions
in a four-dimensional single-particle space?}



\section{Quantum correlations of pure fermionic and bosonic states in higher-dimensional
Hilbert spaces}
\label{Kai}

\subsection{Two-fermion states}
As shown above, correlations of pure states of two fermions with a four-dimensional
single-particle Hilbert space are characterized by the concurrence
${\cal C}(\Ket{w})$. In particular this quantity vanishes if and only if
$\Ket{w}$ has Slater rank one. As soon as larger Hilbert spaces are considered 
there is no straight forward extension of the concurrence. However,
performing the Slater decomposition is always possible and thus correlations of
a two-fermion system can be characterized by the Slater rank. Here
we introduce criteria for pure states in a single-particle space 
of arbitrary (even) dimension $d=2K$ that allow to specify the Slater rank without
explicitly performing the decomposition.

To start with notice that in the case of $2K=4$ the concurrence vanishes
if and only if the general expression (\ref{eqn:asdet})
for the determinant of an antisymmetric $2K\times2K$-matrix is zero. Using this
for general dimension $2K$ together with the block structure of $w$ in the
canonical form it is clear that $\Ket{w}$ has Slater rank $<K$
if and only if 
\begin{equation}
\sum_{i_1,\ldots,i_{2K}=1}^{2K}\e^{i_1\ldots i_{2K}}
w_{i_1i_2}\ldots w_{i_{2K-1}i_{2K}}=0.
\end{equation}

Let us first turn to the case of $2K=6$. Here it only remains to distinguish 
Slater rank one from Slater rank two/three. This can be done by 
\begin{lemma}\label{lem:2ferm6dim}
Let $\Ket{w}=\sum_{i,j=1}^6 w_{ij}f^{\dagger}_i f^{\dagger}_j\Ket{\0}$ be a two-fermion
state in a six-dimensional single-particle space. Then $\Ket{w}$ has
Slater rank one if and only if $\,\forall\,\alpha,\,\beta$, with
$\,1\leq\alpha<\beta\leq 6$,
\begin{equation}
\sum_{i,j,k,l=1}^6\,w_{ij}w_{kl}\,\e^{ijkl\alpha\beta}=0.
\end{equation}
\end{lemma}
The proof is given in the appendix \ref{proofs:fermionic}. 

For $2K>6$ this result can be generalized to classify the Slater rank of an
arbitrary pure state of two fermions. The quantities to be calculated have an analogous
structure: Slater rank two or higher is distinguished from Slater rank one
by contraction of two $w$-matrices with the $\varepsilon$-tensor. Slater rank one
and two can be distinguished from Slater rank three and above by contracting
three $w$-matrices and so on until finally states with maximal Slater rank are 
identified by the full contraction of $K$ matrices. This is collected in the
following lemma (see the appendix for the proof):
\begin{lemma}\label{lem:2FmSrN}
A two-fermion state $
\Ket{w}=\sum_{i,j=1}^{2K}w_{ij}f^{\dagger}_i f^{\dagger}_j\Ket{\0}
\in{\cal A}({\cal H}_{2K}\otimes{\cal H}_{2K})
$
has Slater rank $< N$ if and only if 
$\;\forall\;\alpha_1,\ldots,\alpha_{2(K-N)}$, with $1\,\leq\,\alpha_1\,<\ldots\,<\,\alpha_{2(K-N)}\,\leq\,2K$,
\begin{equation}
\sum_{i_1,\ldots,i_{2N}=1}^{2K}w_{i_1i_2}\ldots w_{i_{2N-1}i_{2N}}
\e^{i_1\ldots i_{2N}\,\alpha_1\ldots\alpha_{2(K-N)}}=0.
\end{equation}
\end{lemma}
\subsection{Three- and multi-fermion states}
Three identical fermions in a $2K$-dimensional single particle space ${\cal H}_{2K}$
live in a total Hilbert space ${\cal A}({\cal H}_{2K}\otimes{\cal H}_{2K}\otimes
{\cal H}_{2K})$. In the usual notation a general pure state is written
\begin{equation}
\Ket{w}=\sum_{i,j,k=1}^{2K}w_{ijk}f_i^{\dagger}f_j^{\dagger}f_k^{\dagger}\Ket{0},
\end{equation}
where $w$ is completely antisymmetric and fulfills the normalization
condition
\begin{equation}
\sum_{i,j,k=1}^{2K} w_{ijk}^*w_{ijk}=\frac16\;.
\end{equation}

Because for fermions every single-particle state can only be occupied once,
a particle hole transformation ${\cal U}_{{\rm ph}}$
(c.f.~eqn.~(\ref{ph1})) allows to treat the cases $d=2K=4$ and $d=5$
more easily: three fermions acting in a four-dimensional single-particle space are
mapped by ${\cal U}_{{\rm ph}}$ onto one fermion (and thus such a state is never correlated)
while a five-dimensional three-fermion state is mapped
onto two fermions in five dimensions. This five-dimensional state can then always
be embedded into a four-dimensional single particle space because the rank of
the coefficient matrix is always even. Thus for $d=4$ a state of three fermions
is always non-correlated, for $d=5$ the concurrence from section \ref{sec:concurrence}
can be used. The lowest non-trivial systems of three fermions act in a
single particle space of dimension $2K=6$.

In the case of distinguishable particles it is well known that a straight forward
generalization of the bi-orthogonal Schmidt decomposition to systems of more than two
parties is not possible \cite{PER:95,THA:98,ACIN:01}. Not every
three-quantum-bit state can be transformed via local unitary transformations into
a sum of tri-orthogonal product states, i.e.
\begin{equation}
\Ket{\psi}=d_0 \Ket{0}_A\otimes\Ket{0}_B\otimes\Ket{0}_C+
d_1 \Ket{1}_A\otimes\Ket{1}_B\otimes\Ket{1}_C
\end{equation}
is not always possible.

Counting of dimensions shows that also for three fermions a canonical form 
consisting of a sum of elementary Slater determinants cannot exist. In the case
of $2K=6$ this would mean to find a unitary transformation ${\cal U}$ of the single
particle space such that
\begin{equation}
w'_{lmn}=\sum_{i,j,k=1}^{2K}w_{ijk}U_{li}U_{mj}U_{nk}=z_1\e^{lmn456}+z_2\e^{123lmn},
\end{equation}
i.e.
$\Ket{w'}\propto z_1\,f_1^{\dagger}f_2^{\dagger}f_3^{\dagger}\Ket{0}+ z_2\,f_4^{\dagger}f_5^{\dagger}f_6^{\dagger}\Ket{0}$,
where by adjusting the phases of the basis states the $z_k$ can be chosen real and non-negative.
In general such a form depends on $\lfloor2K/3\rfloor$ real numbers
($\lfloor x\rfloor$ is the largest integer $\leq x$), while 
a general three-fermion state has
\begin{equation}
2\left[{2K \choose 3}\right]=\frac{(2K)^3-3(2K)^2+4K}{3}
\end{equation}
parameters (including normalization and an overall phase).
A unitary transformation $\cal U$ of the $2K$-dimensional single particle space has
$(2K)^2$ real parameters.
For the case of $2K=6$ this means that a general state is described by $40$ parameters while the desired 
Slater decomposed form only has two parameters. But then the $36$ real numbers that can
be chosen for the unitary transformation are not enough in general to solve
the corresponding set of linear equations.

It is interesting that questions on multi-fermion pure states can be answered
by projecting them onto pure states of fewer particles. For this purpose
let us define an operator $R_{\Ket{a}}$ with
$\Ket{a}=\sum_ia_if_i^{\dagger}\Ket{0}$, which projects the three-fermion
state $\Ket{w}$ on a two-fermion state $\Ket{\hat w}$ supported at most
on the single-particle space orthogonal to $\Ket{a}$ and which is defined through
the following equation:
\begin{equation}\label{eqn:operatorR}
\Ket{\hat w}=R_{\Ket{a}}\Ket{w}=R_{\Ket{a}}
\sum_{i,j,k=1}^{2K}w_{ijk}f_{i}^{\dagger}f_{j}^{\dagger}f_{k}^{\dagger}\Ket{0}
=3 \sum_{i,j=1}^{2K}\,\sum_{k=1}^{2K}w_{ijk}a_k\, f_i^{\dagger}f_j^{\dagger}\Ket{0}.
\end{equation} 

The first thing to be noticed is that a decomposition of $\Ket{w}$
into elementary non-overlapping Slater 
determinants only exists if there is a set
$\left\{\Ket{e^{\alpha}}\right\}_{\alpha=1,\ldots,2K}$, 
$\BraKet{e^{\alpha_1}}{e^{\alpha_2}}=\delta^{\alpha_1\alpha_2}$ and
${\cal H}_{2K}=\mbox{span}\left\{\Ket{e^{\alpha}}\right\}$ such that
$R_{\Ket{e^{\alpha}}}\Ket{w}$ has Slater rank one or is zero and
$R_{\Ket{e^{\alpha}}}\Ket{w}$ is orthogonal to $R_{\Ket{e^{\beta}}}\Ket{w}$ if 
$\alpha\neq\beta$. In the basis where $\Ket{w}$ has the desired block form
this set is given by the set of basis vectors. 

Following the definition of non-correlated states for two-fermion systems a three-fermion state is
non-correlated if and only if it can written as
\begin{equation}
\Ket{w}=f_1^{\dagger}f_2^{\dagger}f_3^{\dagger}\Ket{0}
\end{equation}  
in some basis. This is simply a Slater decomposable state with Slater rank one.
These are also the states called
``separable'' in reference \cite{LZLL:01}. We first remark that here the particle-hole
transformation does not help to identify or even to quantify quantum correlations
as it did via the concurrence in the two-fermion case, because for a three-fermion
state $\Ket{w}$ we have $|\BraKet{\tilde w}{w}|
=|\Bra{w}({\cal U}{\cal K})^{\dagger}\Ket{w}|\equiv0$. This is analogous
to three distinguishable qubits where
$\Bra{\psi}(i\sigma^y_A)\otimes(i\sigma^y_B)\otimes(i\sigma^y_C)){\cal K}\Ket{\psi}\equiv 0$
(c.f.~eqn.~(\ref{twoqubitflip})) \cite{VoWe:00}.

Instead of a direct translation of the concurrence there is a 
necessary and sufficient criterion for three-fermion pure states having Slater rank one, by
using the projection defined above (the proof is given in appendix \ref{proofs:fermionic}:
\begin{lemma}\label{lem:3FmSr1}
A three-fermion state $
\Ket{w}=\sum_{i,j,k=1}^{2K}w_{ijk}f_{i}^{\dagger}f_{j}^{\dagger}f_{k}^{\dagger}\Ket{0}
\in{\cal A}({\cal H}_{2K}\otimes{\cal H}_{2K}\otimes
{\cal H}_{2K})
$
is not correlated (has Slater rank one) if and only if $\forall\Ket{a}\in{\cal H}_{2K}$:
\begin{equation}
\quad R_{\Ket{a}}\Ket{w}=
3 \sum_{i,j=1}^{2K}\,\sum_{k=1}^{2K}w_{ijk}a_k\, f_i^{\dagger}f_j^{\dagger}\Ket{0}
\end{equation}
has Slater rank one or is equal to the $2K$-dimensional zero.
\end{lemma}
This lemma can be combined together with \ref{lem:2FmSrN} (which allowed
to calculate the Slater rank of two-fermion states) to give  
\begin{lemma}\label{lem:3FmSr1v2}
A three-fermion pure state 
$\Ket{w}=\sum_{i,j,k=1}^{2K}w_{ijk}f_{i}^{\dagger}f_{j}^{\dagger}f_{k}^{\dagger}\Ket{0}$
has Slater rank one if and only if
$\forall\;\alpha_1,\ldots,\alpha_{2(K-2)}$, $1\,\leq\,\alpha_1\,<\ldots\,<\,\alpha_{2(K-2)}\,\leq\,2K$,
and $\,\forall\,\vc{a}\in\C^{2K}$,
\begin{equation}
\sum_{i,j,k,l,m,n=1}^{2K}\,w_{ijk}\,a_i\;w_{lmn}\,a_l\,
\e^{jkmn\,\alpha_1\ldots\alpha_{2(K-2)}}=0.
\end{equation}
\end{lemma}
As an example consider
\begin{equation}
\Ket{w}=x\,f_1^{\dagger}f_2^{\dagger}f_3^{\dagger}\Ket{0}+y\,f_3^{\dagger}f_5^{\dagger}f_6^{\dagger}\Ket{0},
\quad w_{ijk}=\frac13\left(x \e^{ijk456}+y\e^{12i4jk}\right)\Ket{0},\quad |x|^2+|y|^2=1.
\label{eqn:ex3FmCorr}
\end{equation}
By lemma \ref{lem:3FmSr1}, $\Ket{w}$ does not have Slater rank one because for
$\vc{a}=(0,0,1,0,0,0)^{T}$
\begin{equation}
R_{\Ket{a}}\Ket{w}\propto x f_1^{\dagger}f_2^{\dagger}\Ket{0}+y f_5^{\dagger}f_6^{\dagger}\Ket{0}
\end{equation}
has Slater rank two. Also evaluation of lemma \ref{lem:3FmSr1v2} shows that $\Ket{w}$
does not have Slater rank one because it gives rise to the five conditions
\begin{equation}
a_3^2\,x\,y=0\quad a_1a_3\,x\,y=0\quad a_2a_3\,x\,y=0
\quad a_3a_5\,x\,y=0\quad a_3a_6\,x\,y=0
\end{equation}
which are obviously violated by some choices of $\vc{a}$ as long as
$x$ and $y$ are non-zero. Notice that also $\Ket{w}$ cannot be written by a
transformation of the single-particle space as a sum of orthogonal Slater determinants as
\begin{equation}
\Ket{w'}\propto z_1\,(f')_1^{\dagger}(f')_2^{\dagger}(f')_3^{\dagger}\Ket{0}+
 z_2\,(f')_4^{\dagger}(f')_5^{\dagger}(f')_6^{\dagger}\Ket{0},\quad{\rm i.e.}\,
w'\propto z_1\,f_1^{\dagger}f_2^{\dagger}f_3^{\dagger}\Ket{0}+z_2\,f_4^{\dagger}f_5^{\dagger}f_6^{\dagger}\Ket{0},
\end{equation}
because $R_{\Ket{b}}\Ket{w'}\neq0$ for all $\Ket{b}$ but
$R_{\Ket{b}}\Ket{w}=0$ for $\Ket{b}=f^{\dagger}\Ket{0}$ (or, equivalently,
$\text{rank}(w')>\text{rank}(w)$).
All the comments and results for three-fermion states can be carried over to 
systems consisting of $N$ identical fermions in a Hilbert space
${\cal A}({\cal H}_{2K}\otimes\ldots\otimes{\cal H}_{2K})$. Here the lowest non-trivial
dimension of the single-particle space is $2N$ since otherwise a treatment through a
particle hole transformation is easier. A general pure state reads
\begin{equation}
\Ket{w}=
\sum_{i_1,\ldots,i_N=1}^{2K}w_{i_1\ldots i_N}f_{i_1}^{\dagger}\ldots f_{i_N}^{\dagger}\Ket{0},
\quad \sum_{i_1,\ldots,i_N=1}^{2K}{w^*}_{i_1\ldots i_N}{w}_{i_1\ldots i_N}=\frac{1}{N!},
\end{equation}
where $w$ is antisymmetric in all indices. Again for $N>2$ a decomposition
into orthogonal elementary Slater determinants is not possible in general, but a
state equivalent to a pure separable state of distinguishable particles
can be said to have Slater rank one if, by a transformation of the single-particle
space, it can be cast into the form
\begin{equation}
\Ket{w'}= \,f_1^{\dagger}\ldots f_N^{\dagger}\Ket{0}.
\end{equation}
This is a Slater decomposable state with Slater rank one and the definition again
coincides with the definition of ``separable'' in reference \cite{LZLL:01}
The operator $R_{\Ket{a}}$ can be generalized to a projector from a system of $N$
to $N-1$ fermions as
\begin{equation}\label{eqn:operatorRforN}
R_{\Ket{a}}\Ket{w}=\Ket{\hat w}\qquad\mbox{with}\quad
\hat w_{i_1\ldots i_{N-1}}=N\sum_{i_N=1}^{2K}w_{i_1\ldots i_N}a_{i_N}
\end{equation}
and it is clear that the necessary criterion for the existence of a decomposition into 
orthogonal Slater determinants given above for three fermions also holds for
more particles. Furthermore lemma \ref{lem:3FmSr1} can be translated as follows:
\begin{lemma}\label{lem:NFmSr1}
A pure $N$-fermion state has Slater rank one if and only if
\begin{equation}
\forall \Ket{a}\in{\cal H}_{2K}:\quad R_{\Ket{a}}\Ket{w}=
N\sum_{i_1,\ldots,i_N=1}^{2K}\,w_{i_1\ldots i_N}a_{i_N}
\, f_{i_1}^{\dagger}\ldots f_{i_{N-1}}^{\dagger}\Ket{0}
\end{equation}
has Slater rank one or is zero.
\end{lemma}
The proof can be found in appendix \ref{proofs:fermionic}.
The lemma is maybe easier to handle if it is given in the following form:
\begin{lemma}\label{lem:NFmSr1v2}
A pure $N$-fermion state has Slater rank one if and only if
$\,\forall\;\alpha_i$, with $1\,\leq\,\alpha_1\,<\ldots\,<\,\alpha_{2(K-2)}\,\leq\,2K$,
and for all $\vc{a}^{\,1},\ldots,\vc{a}^{\,N-2}\in\C^{2K}$,
\begin{equation}
\mathop{\sum_{i_1,\ldots i_N=1}^{2K}}_{j_1,\ldots j_N=1}\,
w_{i_1\ldots i_N}\,a^{1}_{i_1}\ldots a^{N-2}_{i_{N-2}}\;
w_{j_1\ldots j_N}\,a^{1}_{j_1}\ldots a^{N-2}_{j_{N-2}}\,
\e^{i_{N-1}i_Nj_{N-1}j_N\,\alpha_1\ldots\alpha_{2(K-2)}}=0.
\end{equation}
\end{lemma}
Here lemma \ref{lem:NFmSr1} is simply used several times, thereby reducing the
number of fermions in the system to $N-1$, $N-2$ and so on until
lemma \ref{lem:2FmSrN} can be applied to the resulting two-fermion state.

As an example illustrating this lemma and showing that the
straightforward extension of the two-fermion concurrence using the particle-hole
transformation fails, let
\begin{equation}
\Ket{w}=\sum_{i,j,k,l=1}^{2K}w_{ijkl}\fc{i}\fc{j}\fc{k}\fc{l}\Ket{0},\qquad
w_{ijkl}=x \e^{ijkl5678}+ y \e^{ij34kl78}+ z \e^{12ijkl78}.
\end{equation}
To see that this is a correlated state take $\vc{a}=(0,1,0,0,0,0,0,0)^{\T}$ to get
$R_{\Ket{a}}\Ket{w}=x\,\fc{1}\fc{3}\fc{4}\Ket{\0}+y\,\fc{1}\fc{5}\fc{6}\Ket{\0}$, which
after some permutation of the basis is proportional to the three-fermion correlated state
from equation (\ref{eqn:ex3FmCorr}). Thus also $\Ket{w}$ does not have Slater rank one.
Nevertheless
\begin{equation}
\Ket{\widetilde w}=U_{ph}{\cal K}\Ket{w}=
\mathop{\sum_{i,j,k,l=1}^{2K}}_{m,n,o,p=1}\e^{ijklmnop}
{w^*}_{ijkl}\,\fc{m}\fc{n}\fc{o}\fc{p}\Ket{\0}
\end{equation}
and therefore $\BraKet{\widetilde w}{w}=0$ although $\Ket{w}$ contains correlations.



\subsection{Two-boson states}
Next we will apply the concept of assessing correlations by identifying 
the Slater number to bosonic pure states. Given a two-boson state
$\Ket{v}=\sum v_{ij}\bc{i}\bc{j}$ with $v_{ij}=v_{ji}$
a distinction between maximal and non-maximal Slater rank is provided by
the vanishing of $\det v$ in the latter case, i.e.~by considering
the full contraction of $K$ $v$-matrices with two $K$-dimensional
$\e$-tensors. For a finer distinction in the case of a three-dimensional
single-particle space we have
\begin{Lem}\label{lem:2bos3dim}
Let $\Ket{v}=\sum_{i,j=1}^3 v_{ij}\bc{i}\bc{j}\Ket{\0}$
be a two-boson state in a three-dimensional single particle space.
Then $\Ket{v}$ has Slater rank one if and only if $\forall 1\leq\alpha\leq3:$
\begin{equation}\label{eqn:2bos3dim}
\sum_{i,j,k,l=1}^3\,v_{ij}\,v_{kl}\,\e^{ik\alpha}\e^{jl\alpha}=0.
\end{equation}
\end{Lem}

This can be extended to higher-dimensional single-particle spaces essentially
as for fermions. The result is the following lemma which is the bosonic equivalent
to lemma \ref{lem:2FmSrN} (the proof is sketched in appendix \ref{proofs:bosonic}):
\begin{Lem}\label{lem:2bosKdim}
Let $\Ket{v}$ be a two-boson state in $\as{\hil_{K}\otimes\hil_{k}}$.
$\Ket{v}$ has Slater rank $<N$ if and only if
$\,\forall\alpha_1,\ldots,\alpha_{K-N}$, with $1\leq\alpha_1<\alpha_{K-N}\leq K$,
\begin{equation}
\sum_{i_1,\ldots,i_{2N}=1}^K\,v_{i_1i_2}\,v_{i_3i_4}\ldots v_{i_{2N-1}i_{2N}}
\,\e^{i_1i_3\ldots i_{2N-1}\alpha_1\alpha_{K-N}}
\,\e^{i_2i_4\ldots i_{2N}\alpha_1\alpha_{K-N}}=0.
\end{equation}
\end{Lem}
\subsection{Three- and multi-boson states}

For three bosons in a two-dimensional single-particle space a general state $\Ket{v}$ has
$8$ real parameters while a unitary transformation of the single-particle space
has $4$. This shows that, similar to the fermionic case,
a straightforward generalization of the two-boson Slater decomposition is not possible,
i.e. $\Ket{v}$ cannot always be written as
$\Ket{v}=(z_1\,\bc{1}\bc{1}\bc{1}+z_2\,\bc{2}\bc{2}\bc{2})\Ket{\0}$. The same holds for
$N$-boson states with $N\ge 3$.

A special case of $N$-boson states that indeed are Slater decomposable
are non-correlated states that we are going to define now by generalizing the definition
for two-boson systems:
\begin{Def}\label{def:nonCorrNBos}
A pure $N$-boson state $\Ket{v}$ is non-correlated (is Slater decomposable and
has Slater rank one) if in some basis it can be written
\begin{equation}
\Ket{v}=\frac1{\sqrt{N}}\;\bc{1}\ldots\bc{1}\Ket{\0},\qquad{\rm i.e.~}\;
v_{i_1\ldots i_N}=\frac{1}{\sqrt{N}}\;\prod_{j=1}^N\delta_{i_j1}.
\end{equation}
\end{Def}
It is worth noticing that this definition differs slightly from the one in
reference \cite{LZLL:01} where $\Ket{v}$ is also called non-correlated if it can be
written $\Ket{v}=\frac1{\sqrt{N}}\;\bc{1}\ldots\bc{N}\Ket{\0}$ in some basis where
the $bc{i}$ are orthogonal.

If we define an operator $R_{\Ket{a}}$ for $\Ket{a}=\sum_ia_i\bc{i}\Ket{\0}$ 
that projects from a system of $N$ to a system of $N-1$ bosons (c.f.
equations (\ref{eqn:operatorR}) and (\ref{eqn:operatorRforN})),
\begin{equation}
\Ket{\hat v}=R_{\Ket{a}}\Ket{v}=
\sum_{i_1,\ldots,i_N,j=1}^{K}v_{i_1\ldots i_N}a_j\,\ba{j}\bc{i_1}\ldots\bc{i_N}\Ket{\0}
=N\sum_{i_1\ldots i_N=1}^{K}v_{i_1\ldots i_N}a_{i_N}\bc{i_1}\ldots\bc{i_{N-1}}\Ket{\0}
\end{equation}
(here the result is not necessarily only supported on the single-particle
space orthogonal to $\Ket{a}$), 
then an identification of non-correlated states is possible.
\begin{Lem}\label{lem:NbosSr1}
A pure $N$-boson state $\Ket{v}$ has Slater rank one if and only if
the $N-1$-boson state
\begin{equation}
\quad R_{\Ket{a}}\Ket{v}=
N\sum_{i_1,\ldots,i_N=1}^{2K}\,v_{i_1\ldots i_N}a_{i_N}
\, \bc{i_1}\ldots \bc{i_{N-1}}\Ket{0}
\end{equation}
has Slater rank one or is zero $\forall \Ket{a}\in\hil_{2K}$.
\end{Lem}
It is worth to notice that this lemma is also valid
if, instead of the definition \ref{def:nonCorrNBos}, the definition from reference
\cite{LZLL:01} is used. However, concentrating on definition \ref{def:nonCorrNBos},
it is possible to iterate lemma \ref{lem:NbosSr1} until a two-boson state is left
and lemma \ref{lem:2bosKdim} can be applied. Thus we have
\begin{Lem}
An $N$-boson pure state has Slater rank one if and only if
$\forall\alpha_i$ with $1\,\leq\,\alpha_1\,<\ldots\,<\,\alpha_{K-2}\,\leq\,K$,
and for all $\vc{a}^{\,1},\ldots,\vc{a}^{\,N-2}\in\C^{2K}$,
\begin{equation}
\mathop{\sum_{i_1,\ldots i_N=1}^{K}}_{j_1,\ldots j_N=1}
v_{i_1\ldots i_N}a^{(1)}_{i_1}\ldots a^{(N-2)}_{i_{N-2}}\,
v_{j_1\ldots j_N}a^{(1)}_{j_1}\ldots a^{N-2}_{j_{N-2}}\,
\e^{i_{N-1}j_{N-1}\alpha_{1}\ldots\alpha_{K-2}}\,\e^{i_{N}j_{N}\alpha_{1}\ldots\alpha_{K-2}}=0.
\end{equation}
\end{Lem}
%



\section{Quantum correlations in mixed fermionic states}
\label{sumferm}

In this section we repeat for completeness of the present paper 
the main results of Ref. \cite{SCK+:00} concerning  mixed states of two 
identical fermions. Part of these results, concerning the two-fermion system 
with a four-dimensional single particle space has been already discussed 
in Section 2. In this case the question of the Slater rank of a given 
mixed state can be answered exactly. 
Here we concentrate thus on the general case, for which 
no exact answers are known in general.

\subsection{Slater number of mixed fermionic states}

\label{mixedstates}

Let us now generalize the concepts introduced in Section 2
to the case of fermionic mixed
states with a single-particle space of arbitrary dimension.
 To this end, we define formally the Slater number of a mixed state, 
in analogy to the Schmidt number for the case of distinguishable
parties \cite{terhal001,schmidt} : 

\begin{Def}\label{fermdef1}
\rm  {Consider a  density matrix
$\rho$ of a two-fermion  system,  and all its possible convex
decompositions in terms of pure states, i.e.
$\rho=\sum_i p_i \Ket\psi_i^{r_i}\Bra\psi_i^{r_i}$,
where $r_i$ denotes the Slater rank of $\Ket\psi^{r_i}_i$; the Slater  number
of $\rho$, $k$, is defined as $k=\min\{r_{\max}\}$,
where ${r}_{\max}$ is the maximum Slater rank within a decomposition,
and the minimum is taken over all decompositions.}
\end{Def}
In other words, there exists a decomposition of $\rho$ that 
uses pure states with Slater rank not exceeding $k$, and there is no 
construction that only uses pure states with Slater rank smaller than $k$.

Many of the results concerning Schmidt numbers can be  
transferred directly to the Slater number.  
For instance, let us denote the whole space of density matrices in
${\cal A}({\cal C}^{2K}\otimes{\cal C}^{2K})$ by
$Sl_K$, and the set of density matrices that have Slater number $k$ or
less,  by $Sl_k$. $Sl_k$ is a convex compact subset of
$Sl_K$.    A state from $Sl_k$ will be called a state of
(Slater) class $k$. Sets of increasing Slater number are embedded into
each other, i.e. $Sl_1\subset Sl_2\subset...Sl_k ...\subset Sl_K$.
In particular, $Sl_1$ is the set of states
that can be written as a convex combination of elementary Slater
determinants; $Sl_2$ is the set of states of Slater number 2, 
i.e. those that require at least one pure state of Slater rank 2 for
their formation, etc. 

The determination of the Slater number of a given state is in general 
a very difficult task.
Similarly, however, as in the case of separability of mixed states of
two qubits (i.e. states in 
${\cal C}^{2}\otimes{\cal C}^{2}$), and one qubit--one qutrit (i.e.
states in ${\cal C}^{2}\otimes{\cal C}^{3}$) \cite{horo96}, the
situation is particularly simple in the case of small $K$.  For $K=1$
there exists only one state (a singlet). For $K=2$
we have  presented in section 2
 a necessary and sufficient condition for a given mixed 
state to have a Slater number of one. 
We have also defined the Slater correlation measure in analogy to 
 Wootters' result \cite{Woo:98},
relating separability to a quantity called
concurrence, and to eigenvalues of a certain matrix. 
One should note, however, that in the considered case  of
fermionic states there exists no simple analogy of the partial
transposition, which is essential for the theory of entangled states of distinguishable
particles. 
In fact, the Peres--Horodecki criterion \cite{horo96,peres96} in  $2\times
2$ and $2\times 3$ spaces says that a state is separable if and only if its partial
transpose is positive semidefinite.  The results of section 2
can be formulated as the following Theorem:    

\begin{Thm}\label{fermthm1}
{\rm Let the mixed state acting in ${\cal A}({\cal C}^{4}\otimes{\cal
C}^{4})$ have a spectral decomposition
$\rho=\sum_{i=1}^r|\Psi_i\rangle\langle\Psi_i|$, where $r$ is the rank of
$\rho$, and the eigenvectors $|\Psi_i\rangle$ belonging to nonzero
eigenvalues $\lambda_i$ 
are normalized as
$\langle\Psi_i|\Psi_j\rangle=\lambda_i\delta_{ij}$. 
Let $|\Psi_i\rangle=\sum_{a,b}w^i_{ab}
f^{\dag}_af^{\dag}_b|0\rangle$ in some basis, and define the complex 
symmetric $r\times r$ matrix $C$ by
\begin{equation}
C_{ij}=\sum_{abcd}\varepsilon^{abcd}w^i_{ab}w^j_{cd},
\label{cma}
\end{equation}
which   can be represented using a unitary matrix as 
$C=UC_dU^T$, with $C_d={\rm diag}[c_1,c_2,\ldots,c_r]$ diagonal and 
$|c_1|\ge|c_2|\ge \dots\ge |c_r|$. The state $\rho$ has Slater number 1
if and only if 
\begin{equation}
|c_1|\le \sum_{i=2}^r |c_i|.
\end{equation}}
\end{Thm}

The above Theorem is thus 
an analogue of the Peres-Horodecki-Wootters result for
two-fermion systems having a single particle space of dimension
$2K\le 4$.  The
situation is much more complicated, when we go to
$K>2$; this is similar to the case of the separability problem in ${\cal
C}^M\otimes{\cal C}^N$ with $MN>6$. These issues are investigated in the next 
subsection.
 
At this point it is worth stressing that our formalism can also be used for odd
dimensions, i.e.\ formally for half-integer $K$. Here the situation is again
particularly simple for $K=1.5$ ($2K=3$). All states have obviously the Slater
number one and can thus be written as a convex sum of projectors onto states
equivalent to the singlet state $\Ket{0}\Ket{1}-\Ket{1}\Ket{0}$, which implies
that the entanglement of formation of any antisymmetric mixed state (treated as a state of two distinguishable particles) is $1$ ebit.
This has been observed recently by Vidal, D{\"u}r and Cirac \cite{vidal:2002},
who showed that also the entanglement cost of every such
antisymmetric state is $1$ ebit which implies additivity of the entanglement of formation.


\subsection{Fermionic Slater witnesses}
\label{witnesses}

We now investigate fermionic systems with single-particle Hilbert
spaces of dimension $2K>4$. In this case, a full and explicit 
characterization of pure and mixed state 
quantum correlations, such as given above for the two-fermion system with
$K=2$, is apparently not possible.
Therefore one has to
formulate other methods to investigate the Slater number of a given state.
We can, however, as in Ref. \cite{SCK+:00} follow here the lines 
of the papers that we have
written on entanglement witnesses\cite{opti,mapy}, and Schmidt number
witnesses \cite{schmidt}. In the following we skip the proofs, which are discussed in detail in Ref. \cite{SCK+:00}.

In order to determine the Slater number of a fermionic density matrix
$\rho$ we note that due to the fact that the sets $Sl_k$ are convex and
compact,  any  density matrix of class $k$  can be decomposed into
a  convex combination of a density matrix of class $k-1$,  and  a
remainder $\delta$\cite{Le98}:

\begin{Prop}\label{fermprop1}
\rm { Any state of class $k$, $\rho_k$,  can be
written as a convex combination of a density matrix of class $k-1$ and
a so-called $k-$edge state $\delta$:
\begin{equation}
\rho_k  =(1-p) \rho_{k-1} + p \delta, \;\;\;\, 1\ge p> 0 ,
\label{decom}
\end{equation}
where the edge state $\delta$ has Slater  number $\ge k$.}
\end{Prop}

The decomposition (\ref{decom}) is obtained by
subtracting projectors onto pure states of Slater rank smaller than
$k$, $P=\Ket{\psi^{<k}}\Bra{\psi^{<k}}$
such that $\rho_k-\lambda P\ge 0$. Here $\Ket{\psi^{<k}}$ stands for 
pure states of Slater rank $r <k$.
Denoting by $K(\rho)$, $R(\rho)$, and
$r(\rho)$ the kernel, range, and rank of $\rho$, respectively,   
we observe that $\rho'\propto \rho-\lambda {\Ket\psi}{\Bra\psi}$  
is non negative if and only if ${\Ket\psi}\in R(\rho)$ 
and $\lambda\le {\Bra\psi}\rho^{-1}{\Ket\psi}^{-1}$ (see \cite{Le98}).
The idea behind this decomposition is that the edge state $\delta$
which has generically lower rank contains all the information
concerning the Slater number $k$ of
the density matrix $\rho_k$. 

As in the case of the Schmidt number, there 
is an optimal decomposition of the form (\ref{decom}) with $p$ minimal.
Alternatively, restricting ourselves to decompositions
$\rho_k=\sum_i p_i \Ket\psi_i^{r_i}\Bra\psi_i^{r_i}$
with all $r_i\le k$, we can always find a decomposition
of the form (\ref{decom}) with  $\delta\in Sl_k$. We define below more
precisely what an edge state is.

\begin{Def}\label{fermdef2}
\rm {A fermionic $k$-edge state $\delta$
is a state such that
$\delta-\varepsilon {\Ket{\psi^{<k}}}{\Bra{\psi^{<k}}}$ is not positive,
for any $\varepsilon>0$ and $\Ket{\psi^{<k}}$.} 
\end{Def}

The edge states are thus  characterized by the following 

\begin{Crit}\label{fermcrit1} \rm{A mixed state
$\delta$ is a $k$-edge state if and only if there exists no
 $\Ket{\psi^{<k}}$ such that  $\Ket{\psi^{<k}}\in R(\delta)$.}
\end{Crit}

Now we are in the position of defining 
 a fermionic $k$-Slater witness ($k$-SlW, $k\ge 2$):  
\begin{Def}\label{fermdef3}
\rm{ A hermitian operator $W$ is a fermionic Slater witness (SlW) of class $k$ 
if and only if Tr$(W\sigma)\ge 0$ for all $\sigma\in Sl_{k-1}$, and there exists 
at least one fermionic $\rho \in Sl_k$ such that Tr$(W\rho)< 0$.}
\end{Def}

It is straightforward to see that every SlW that detects $\rho$
given by (\ref{decom}) also detects the edge state $\delta$,
since if Tr$(W\rho)<0$  then necessarily Tr$(W\delta)<0$, too. 
Thus, the knowledge of  all SlW's of $k$-edge states fully characterizes 
all $\rho\in Sl_{k}$. Below, we show how to
construct for any edge state a SlW which detects it.
Most of the technical proofs used to construct and optimize
fermionic Slater witnesses are practically equivalent to those presented in 
Ref.\cite{opti} for entanglement witnesses.

All the operators we consider in this subsection  act in ${\cal A}({\cal
C}^{2K}\otimes{\cal C}^{2K})$. Let $\delta$ be a $k$-edge state,
$C$ an arbitrary positive operator such that ${\rm Tr}(\delta
C)>0$, and $P$ a positive operator whose range fulfills
$R(P)=K(\delta)$. We define
$\label{epsilon1} 
\varepsilon\equiv  \inf_{\Ket{\psi^{<k}}}{\Bra{\psi^{<k}}}P{\Ket{\psi^{<k}}}$ and 
$ c\equiv \sup {\Bra\psi}C{\Ket{\psi}}$. 
Note that $c>0$ by construction and $\varepsilon > 0$, 
because $R(P)= K(\delta)$ and 
therefore, since $R(\delta)$ does not contain any ${\Ket{\psi^{<k}}}$ 
by the definition of edge state, $K(P)$ cannot contain any  
${\Ket{\psi^{<k}}}$ either. This implies:
\begin{lemma}\label{eqn:slaterwferm}
{\rm
Given a fermionic $k$-edge state $\delta$, then
\begin{equation}
W = P-{\varepsilon \over c}C
\end{equation}
is a fermionic $k$-SlW which detects $\delta$.} 
\end{lemma}
The simplest choice of $P$ and $C$ consists in taking 
projections onto $K(\delta)$ and the identity
operator on the antisymmetric space ${\Eins}_a$, respectively. As we will see
below, this choice provides us with a canonical form of a fermionic $k$-SlW. 

\begin{Prop} \label{fermprop2}
{\rm
Any fermionic Slater witness can be written in the {\it canonical} form:
\begin{equation}
W=\tilde{W}-\varepsilon {\Eins}_a\ ,
\end{equation}
such that $R(\tilde W)= K(\delta)$, where $\delta$ is a
$k$-edge state and $0<\varepsilon\le {\rm inf}_{|\psi^{<k}\rangle
}{\Bra{\psi}}\tilde{W}{\Ket{\psi}}$}.
\label{prop2}
\end{Prop}

It is useful to consider also
\begin{Def}
{\rm A $k$-Slater witness $W$ is {\it tangent} to $Sl_{k-1}$ at $\rho$ 
if $\exists$ a state $\rho\in  Sl_{k-1}$ such that 
Tr$(W\rho)=0$}.
\end{Def}
Similar methods to the ones used in ref.~\cite{SCK+:00} allow to prove the Proposition \ref{prop2} and  to observe that
\newtheorem{guess5}{Observation}
\begin{Observ}
\rm{The fermionic state $\rho$ is of Slater class $k-1$ if and only if
for all fermionic $k$-SW's tangent to $Sl_{k-1}$, Tr$(W\rho)\ge 0$.}
\end{Observ}

\subsubsection{Optimal fermionic Slater witnesses}
\label{optwitness}
We will now discuss the optimization of a fermionic Slater witness.
As proposed in \cite{opti} and \cite{schmidt} an entanglement
witness (Schmidt witness) W is optimal  if there exists no other witness that
detects more states than it.  The same definition can be applied to
fermionic Slater witnesses.  We say that a fermionic $k-$Slater witness 
$W_2$ is  finer  
than   a fermionic $k-$Slater witness $W_1$, if 
$W_2$ detects more fermionic states than $W_1$. Analogously, we define a
$k-$Slater witness  $W$ to be optimal when 
there exists no finer witness than itself. 
Let us define the set of ${\Ket\psi}$ 
pure states of Slater rank $k-1$ 
for which the expectation value of the
$k$-Slater witness $W$ vanishes:
\begin{equation}
T_{W}=\{ \Ket\psi\ \st \ \Bra\psi W \Ket\psi=0 \}\ ,
\end{equation}
i.e. the set of pure tangent states of Slater rank $<k$.
$W$ is an optimal $k$-SlW if and only if $W-\varepsilon P$ is not a $k$-SlW,
for any positive operator $P$. If the set
$T_{W}$ spans the whole Hilbert space ${\cal A}({\cal
C}^{2K}\otimes{\cal C}^{2K})$, then $W$ is an optimal $k$-SlW.
If $T_{W}$ does not span ${\cal A}({\cal
C}^{2K}\otimes{\cal C}^{2K})$, 
then we can optimize the witness by subtracting from 
it a positive operator
$P$, such that $PT_{W}=0$. For example,  
for Slater witnesses of class 2 this is possible, provided that 
$\inf_{|e\rangle \in{\cal
C}^{2K}}[P_{e}^{-1/2}W_{e}P_{e}^{-1/2}]_{\rm min}>0$. 
Here for any $X$ acting on ${\cal A}({\cal C}^{2K}\otimes{\cal C}^{2K})$
we define
\begin{eqnarray}
X_{e} & = & \Big[\langle e,.|X|e,.\rangle 
-\langle e,.|X|.,e\rangle\nonumber\\
& - & \langle ., e|X|e,.\rangle + \langle .,e|X|.,e\rangle\Big]\,, 
\end{eqnarray}
as an operator acting in  ${\cal C}^{2K}$, and $[X]_{\rm
min}$ denotes its minimal eigenvalue (see\cite{opti}).
An example of an optimal witness of Slater  number $k$ 
in ${\cal A}({\cal
C}^{2K}\otimes{\cal C}^{2K})$ is given by
\begin{equation}
W={\Eins}_a-\frac{K}{k-1}{\cal P} \;,
\label{example}
\end{equation}
where ${\cal P}$ is a projector onto a ``maximally correlated state'',
$|\Psi\rangle=\frac{1}{\sqrt{K}}
\sum_{i=1}^{K}f_{2i}^{\dag}f_{2i+1}^{\dag}|0\rangle$.
The reader can easily check that the above witness operator has mean value zero
in the states  $f_{2i}^{\dag}f_{2i+1}^{\dag}|0\rangle$ for 
$i=1,2$, but also
for all states of the form $g_{1}^{\dag}g_{2}^{\dag}|0\rangle$ where
\begin{eqnarray}
g_1^{\dag} & = & f_{1}^{\dag}e^{i\varphi_{11}}
+f_{2}^{\dag}e^{i\varphi_{12}}\nonumber\\
 & & +f_{3}^{\dag}e^{i\varphi_{21}}
+f_{4}^{\dag}e^{i\varphi_{22}},\\
g_2^{\dag} & = & -f_{1}^{\dag}e^{-i\varphi_{12}}
+f_{2}^{\dag}e^{-i\varphi_{11}}\nonumber\\
 & & -f_{3}^{\dag}e^{-i\varphi_{22}}
+f_{4}^{\dag}e^{-i\varphi_{21}},
\end{eqnarray}
for arbitrary $\varphi_{ij}$, $i,j=1,2$. The set $T_W$ spans  in this
case the whole Hilbert space ${\cal A}({\cal
C}^{2K}\otimes{\cal C}^{2K})$, {\it ergo} $W$ is optimal.

\subsubsection{Fermionic Slater witnesses and positive maps}

It is interesting also to 
consider positive maps associated with fermionic Slater witnesses via the Jamio\l
kowski isomorphism \cite{Jam:72}. Such maps employ $W$ acting in ${\cal
H}_A\otimes{\cal H_B}={\cal C}^{2K}\otimes{\cal C}^{2K}$, and transform a
state $\rho$ acting in
${\cal H}_A\otimes{\cal H_C}={\cal C}^{2K}\otimes{\cal C}^{2K}$ into
another state acting in
${\cal H}_B\otimes{\cal H_C}={\cal C}^{2K}\otimes{\cal C}^{2K}$,
via the relation 
$M(\rho)={\rm Tr}_A(W\rho^{T_A})$. Obviously, such maps are positive on
separable states: When $\rho$ is separable, then for any $|\Psi\rangle\in
{\cal H}_B\otimes{\cal H_C}$, the mean value of
$\langle\Psi|M(\rho)|\Psi\rangle$, becomes  a convex sum of mean values 
of $W$ in some product states $|e,f\rangle\in{\cal
H}_A\otimes{\cal H_B}$. Since $W$ acts in fact in the antisymmetric space,
we can antisymmetrized these states, i.e. $|e,f\rangle\to (|e,f\rangle-|f,e\rangle)$.
Such antisymmetric states have, however, Slater rank 1, and all SlW of
class $k\ge 2$ have thus positive  mean value in those states. This class
of positive maps is quite different from the ones considered in Refs.
\cite{opti,mapy}; they provide thus an interesting
class of necessary separability conditions. 
The particular map associated with the witness (\ref{example}) is, however, 
decomposable, i.e. it is a sum of a completely positive map and another completely
positive map composed with the transposition.  This follows from the fact that $W^{\T_A}=Q$
is a positive operator and from $M(\rho)=\Tr_A(W\rho^{\T_A})=\Tr_A(Q\rho)=M_Q(\rho^{\T_A})
=(M_Q\circ \T_A)(\rho)$ where $M_Q(X)=\Tr_A(QX^{\T_A})$ is a completely positive
map because $Q$ is a positive operator.


\section{Quantum correlations in mixed bosonic states}
\label{bosons}

Here we present new results that generalize the
previous section to the case of bosons. 
The case of a two-boson system with a two-dimensional single-particle
space, where the Slater number of a given mixed state can be calculated directly,
has already been discussed (c.f.~section \ref{analogs}). The concern of
this section is thus a single particle Hilbert space ${\cal C}^K$ with $K>2$.
We discuss also the relation of the Slater number and positive definiteness
of the partial transpose.

\subsection{Slater number of mixed bosonic states}

\label{bos:mixedstates}

Following the results from the previous section we may define formally
the Slater number $k$ of a bosonic mixed state.
\begin{bosdef1}
\rm  {Consider a  density matrix
$\rho$ of a two-boson  system,  and all its possible convex
decompositions in terms of pure states, i.e.
$\rho=\sum_i p_i \Ket\psi_i^{r_i}\Bra\psi_i^{r_i}$,
where $r_i$ denotes the Slater rank of $\Ket\psi^{r_i}_i$; the Slater  number
of $\rho$, $k$, is defined as $k=\min\{r_{\max}\}$,
where ${r}_{\max}$ is the maximum Slater rank within a decomposition,
and the minimum is taken over all decompositions.}
\end{bosdef1}

Most of the results concerning Slater numbers for fermions can be  
transferred directly to the Slater number for bosons.  
If we denote the whole space of density matrices in
${\cal S}({\cal C}^{2K}\otimes{\cal C}^{2K})$ by
$Sl_K$, and the set of density matrices that have Slater number $k$ or
less,  by $Sl_k$, then $Sl_k$ as in previous section 
is a convex compact subset of
$Sl_K$;   a state from $Sl_k$ will be called a state of
(Slater) class $k$. Sets of increasing Slater number are embedded into
each other, i.e. $Sl_1\subset Sl_2\subset...Sl_k ...\subset Sl_K$.
In particular, $Sl_1$ is the set of states
that can be written as a convex combination of elementary Slater
permanents.
There exists, however, one very basic difference between bosonic 
and fermionic systems; in the latter case, namely, 
 the concept of partial transposition seems to be completely 
inappropriate. On the other hand, since in bosonic systems 
$Sl_1$ is the set of separable states, partial transposition provides 
here a necessary condition for a state to belong to $Sl_1$. For bosonic 
systems it is thus reasonable to ask questions of the sort: Do 
bosonic    entangled states with positive partial transpose (PPTES) exist?

For the case of $K=2$ a necessary and sufficient condition for a mixed
state to have a Slater number one is given either by the Peres-Horodecki criterion,
i.e.~ by the positivity of the partial transpose, or by the Wootters approach
as discussed in section \ref{analogs}. The latter result we
state again in the following theorem: 

\begin{bosthm1}
{\rm Let the bosonic mixed state acting in ${\cal S}({\cal C}^{2}\otimes{\cal
C}^{2})$ have a spectral decomposition
$\rho=\sum_{i=1}^r|\Psi_i\rangle\langle\Psi_i|$, where $r$ is the rank of
$\rho$, and the eigenvectors $|\Psi_i\rangle$ belonging to nonzero
eigenvalues $\lambda_i$ 
are normalized as
$\langle\Psi_i|\Psi_j\rangle=\lambda_i\delta_{ij}$. 
Let $|\Psi_i\rangle=\sum_{a,b}v^i_{ab}
b^{\dag}_ab^{\dag}_b|\Omega\rangle$ in some basis, and define the complex 
symmetric $r\times r$ matrix $C$ by
\begin{equation}
C_{ij}=\sum_{abcd}\varepsilon^{ac}\varepsilon^{bd}v^i_{ab}v^j_{cd},
\label{bos:cma1b}
\end{equation}
($\varepsilon^{01}=-\varepsilon^{10}=1$, and
 $\varepsilon^{aa}=0$ otherwise) which   can be represented using a unitary matrix as 
$C=UC_dU^T$, with $C_d={\rm diag}[c_1,c_2,\ldots,c_r]$ diagonal and 
$|c_1|\ge|c_2|\ge \dots\ge |c_r|$. The state $\rho$ has Slater number 1
if and only if 
\begin{equation}
|c_1|\le \sum_{i=2}^r |c_i|.
\end{equation}}
\end{bosthm1}


\subsection{Bosonic Slater witnesses}
\label{bos:witnesses}

We now turn to the more difficult case of $K>2$ where as for fermions no
complete characterization of pure and mixed state quantum correlations can be given.
Again the concept of witnesses turns out to be useful here and we can directly translate
most of the fermionic results from section \ref{witnesses}. We will only list the theorems here and skip the
proofs which are discussed in detail in \cite{SCK+:00}.

\begin{bosprop1}
\rm { Any bosonic state of class $k$, $\rho_k$,  can be
written as a convex combination of a density matrix of class $k-1$ and
a so-called bosonic $k-$edge state $\delta$:
\begin{equation}
\rho_k  =(1-p) \rho_{k-1} + p \delta, \;\;\;\, 1\ge p> 0 ,
\label{bos:decom}
\end{equation}
where the edge state $\delta$ has Slater  number $\ge k$.}
\end{bosprop1}

This again follows from the convexity and compactness of the sets $Sl_k$.
As in the case of fermions, there 
is an optimal decomposition of the form (\ref{bos:decom}) with $p$ minimal.
Alternatively, restricting ourselves to decompositions
$\rho_k=\sum_i p_i \Ket\psi_i^{r_i}\Bra\psi_i^{r_i}$
with all $r_i\le k$, we can always find a decomposition
of the form (\ref{bos:decom}) with  $\delta\in Sl_k$.
This allows to define $k$-edge states and $k$-Slater witnesses as in fermionic case.
\begin{bosdef2}
\rm {A bosonic $k$-edge state $\delta$
is a bosonic state such that
$\delta-\varepsilon {\Ket{\psi^{<k}}}{\Bra{\psi^{<k}}}$ is not positive,
for any $\varepsilon>0$ and $\Ket{\psi^{<k}}$.} 
\end{bosdef2}
\begin{boscrit1} \rm{A bosonic mixed state
$\delta$ is a bosonic $k$-edge state if and only if there exists no
 $\Ket{\psi^{<k}}$ such that  $\Ket{\psi^{<k}}\in R(\delta)$.}
\end{boscrit1}  
\begin{bosdef3}
\rm{ A hermitian operator $W$ is a bosonic Slater witness of class $k$ ($k$-SlW, $k\ge 2$) 
if and only if Tr$(W\sigma)\ge 0$ for all $\sigma\in Sl_{k-1}$, and there exists 
at least one bosonic $\rho \in Sl_k$ such that Tr$(W\rho)< 0$.}
\end{bosdef3}

With the same definitions as in the paragraph before 
lemma \ref{eqn:slaterwferm} in section \ref{witnesses}
(but now all the operators act in the symmetric space 
${\cal S}({\cal C}^{K}\otimes{\cal C}^{K})$) we have
\begin{boslem1}
{\rm
Given a bosonic $k$-edge state $\delta$, then
\begin{equation}
W = P-{\varepsilon \over c}C
\end{equation}
is a bosonic $k$-SlW which detects $\delta$.} 
\end{boslem1}
Similarly as in section \ref{witnesses}, the simplest choice of 
$P$ and $C$ consists in taking 
projections onto $K(\delta)$ and the identity
operator on the symmetric space ${\Eins}_s$, respectively.
\begin{bosprop2}
{\rm
Any bosonic Slater witness can be written in the {\it canonical} form:
\begin{equation}
W=\tilde{W}-\varepsilon {\Eins}_s\ ,
\end{equation}
such that $R(\tilde W)= K(\delta)$, where $\delta$ is a
bosonic $k$-edge state and $0<\varepsilon\le {\rm inf}_{|\psi^{<k}\rangle}{\Bra{\psi}}\tilde{W}{\Ket{\psi}}$}.
\label{bos:prop2}
\end{bosprop2}

\subsubsection{Optimal bosonic Slater witnesses}

We will now discuss the optimization of  bosonic Slater witnesses,
applying the same definitions 
of finer and optimal as for fermions. Again we have that if the set $T_W$
of pure tangent states of Slater rank $<k$,
\begin{equation}
T_{W}=\{ \Ket\psi\  \st\  \Bra\psi W \Ket\psi=0 \}\ ,
\end{equation}
spans the complete Hilbert space ${\cal S}({\cal C}^{K}\otimes{\cal C}^{K})$,
then $W$ is an optimal $k$-SlW. Otherwise 
we can optimize the witness by subtracting from it a positive operator
$P$, such that $PT_{W}=0$. For example,  
for Slater witnesses of class 2 this is possible provided that 
$\inf_{|e\rangle \in{\cal
C}^{K}}[P_{e}^{-1/2}W_{e}P_{e}^{-1/2}]_{\rm min}>0$. 
Here for any $X$ acting on ${\cal S}({\cal C}^{K}\otimes{\cal C}^{K})$
we define
\begin{eqnarray}
X_{e} & = & \Big[\langle e,.|X|e,.\rangle 
+\langle e,.|X|.,e\rangle\nonumber\\
& + & \langle ., e|X|e,.\rangle + \langle .,e|X|.,e\rangle\Big]\,, 
\end{eqnarray}
as an operator acting in  ${\cal C}^{K}$, and $[X]_{\rm
min}$ denotes its minimal eigenvalue (compare \cite{opti}).
An example of an optimal bosonic witness of Slater  number $k$ 
in ${\cal S}({\cal
C}^{K}\otimes{\cal C}^{K})$ is given by
\begin{equation}
W={\Eins}_s-\frac{K}{k-1}{\cal P} \;,
\label{bos:example}
\end{equation}
where ${\cal P}$ is a projector onto a ``maximally correlated state'',
$|\Psi\rangle=\frac{1}{\sqrt{K}}
\sum_{a=1}^{K}b_{a}^{\dag}b_{a}^{\dag}|\Omega\rangle$. That this witness is optimal
can be seen as in section \ref{optwitness}.

Finally we only want to mention that it is again possible to relate
positive maps to bosonic Slater witnesses, leading to another interesting class
of necessary separability conditions. As for fermions (c.f.~section \ref{optwitness})
the map associated with the witness (\ref{bos:example}) is decomposable.

%
%
\subsection{Bosonic entangled states and positive partial transpose}

In the case of distinguishable particles there are strong indications
that the Schmidt rank of entangled states with the PPT property is not
maximal, and rather low. This has been conjectured for the states in a
$3\times 3$-dimensional space in Ref. \cite{schmidt}, and for the states
in a
$N\times N$-dimensional space in Ref. \cite{rare}. We expect that similar
situations take place for the Slater rank of bosonic states. In fact we
have not been able to find a two-boson mixed state
that would have the PPT property and would be nevertheless entangled, i.e.
it would have Slater rank strictly greater than 1. For bosonic 3-qubit states
as well as for bosonic $N$-qubit states of rank $\leq N$ we were even able to
prove that such states do not exist.

Our findings are summarized in the following two theorems. The first one
deals with the case of $K=3$, i.e. three-dimensional single-particle
space, but can be  easily generalized to arbitrary $K$. The second one
deals with bosonic systems of $N$ qubits, i.e. N bosons in a two-dimensional
single particle space. 

\begin{Thm}
{\rm Let a bosonic mixed state $\rho$ acting in ${\cal S}({\cal
C}^{3}\otimes{\cal C}^{3})$, which we treat as a symmetrized space of
Alice and Bob, have a positive partial transpose with respect to, say,
Alice. Then
\begin{enumerate}
\item\label{pptthm1.1} if the rank of $\rho$, $r(\rho)=3$, then the Slater rank of $\rho$,
$Sl(\rho)=1$, i.e. $\rho$ is separable.
\item\label{pptthm1.2} if the rank of $\rho$, $r(\rho)=4$, then generically there
exist exactly 4 product vectors of the form $|e_i,e_i\rangle$ in the range
of $\rho$, and the Slater rank of
$\rho$, $Sl(\rho)=1$, i.e. $\rho$  is separable, provided $\langle
e_i,e_i|\rho^{-1}|e_j,e_j\rangle=p_i\delta_{ij}$, where the inverse is
taken on the range. In such case 
$\rho=\sum_{i=1}^4p_i|
e_i,e_i\rangle\langle e_i,e_i|$.
\end{enumerate}
}
\end{Thm}

\noindent {\it Proof.} This theorem is a simple generalization of the 
result of Ref. \cite{nxm}. In particular from  Theorem 1 of that paper
follows that any (symmetric or not) PPT state of rank $3$ is separable,
which proves \ref{pptthm1.1}. To prove \ref{pptthm1.2}, we observe that the vector $|e,e\rangle$
is in the range of $\rho$ if it is orthogonal to vectors which span 
the kernel of $\rho$. There are two such vectors $\Ket{\phi_j}$, $j=1,2$, since
the whole space 
${\cal S}({\cal
C}^{3}\otimes{\cal C}^{3})$ is 6 dimensional. Since in a convenient basis
 up to normalization we can write $|e\rangle=(1,z_1,z_2)$, the
equations $\langle \phi_j|e,e\rangle=0$ constitute a system of two
quadratic polynomials in $z_1$ and $z_2$. Such a system of equations
has generically 4 solutions $|e_i,e_i\rangle$, $i=1,\ldots,4$. From that it
follows that $\rho$ is separable if and only if $\rho=\sum_{i=1}^4p_i|
e_i,e_i\rangle\langle e_i,e_i|$. Using this equation, we observe that
if $\rho^{-1}|e_i,e_i\rangle=|v\rangle$, then
$|e_i,e_i\rangle=\rho|v\rangle$ and from the linear independence of
$|e_i,e_i\rangle$ we get that $\langle e_j,e_j|v\rangle=0$ for $j\ne i$, 
and $\langle e_i,e_i|v\rangle=1$. This implies that $\langle
e_i,e_i|\rho^{-1}|e_j,e_j\rangle=p_i\delta_{ij}$. 

\begin{Thm}
{\rm Let a bosonic mixed state $\rho$ acting in the symmetric $N$-qubit space 
${\cal S}({\cal C}^{2}\otimes\ldots \otimes{\cal C}^{2})$, which we treat
as a symmetrized space of Alice, Bob, Charlie etc., have a positive partial
transpose with respect to, say, Alice.
For $N>3$ let $r(\rho)\le N$. Then $\rho$ is separable.}
\end{Thm}

\noindent{\it Proof.}   Let us consider first the case $N=3$. The state
$\rho$ is bosonic, i.e. has a maximal rank 4, which is the dimension of
the space of totally symmetric vectors. Suppose that the state $\rho$ is,
however, also a PPT state with respect to Alice. We can treat it then 
as a PPT state in a $2\times 4$ dimensional space of Alice and Bob and
Charlie grouped together. Such a state of rank 4 is necessarily separable
and can be represented as $\rho=\sum_{i=1}^4p_i|
e_i,\Psi_i\rangle\langle e_i,\Psi_i|$, where the projectors onto
the vectors
$|\Psi_i\rangle$ are in general entangled pure states of Bob and
Charlie \cite{2xn}. Since they belong to the range of the reduced density
matrix
${\rm Tr}_A\rho$, they must belong to ${\cal S}({\cal
C}^{2}\otimes{\cal C}^{2})$. Obviously, the vectors
$|e_i,\Psi_i\rangle$ belong to the range of $\rho$, and thus to the
symmetric space ${\cal S}({\cal C}^{2}\otimes{\cal C}^{2}\otimes{\cal
C}^{2})$. Direct inspection shows that the only vectors of this
form are
$|e_i,e_i,e_i\rangle$: writing the Schmidt decomposition $|\Psi_i\rangle=
a|00\rangle+b|11\rangle$, and expanding $|e_i\rangle=\alpha|0\rangle
+\beta|1\rangle$ in the same basis, we immediately see that both $\alpha b$
and $\beta a$ must be equal to zero.

The proof for higher $N$ is analogous. We assume that $\rho$ has
rank $\le N$ and a positive partial transpose with
respect to Alice. Since $\rho$, regarded as a matrix acting in a $2\times
2^{N-1}$-dimensional space, is supported  in reality in ${\cal
C}^{2}\otimes{\cal S}(({\cal C}^{2})^{\otimes N-1})$, i.e. in a
$2\times N$-dimensional space, the results of Ref. \cite{2xn}
 hold and $\rho$ must be
separable with respect to the partition Alice -- the rest, and 
therefore have a form $\rho=\sum_{i=1}^Np_i|
e_i,\Psi_i\rangle\langle e_i,\Psi_i|$. But
then, using the analogous argumentation as above we see that
$|e_i,\Psi_i\rangle=|e_i, e_i, \dots,e_i\rangle$.



\section{Conclusions}
\label{concl}

In conclusion, we have studied and characterized quantum correlations in
systems of (i) two distinguishable particles, (ii) two fermions and (iii) two bosons. As an equivalent
to the well-known Schmidt decomposition and rank for distinguishable particles we
introduced the concepts of Slater decomposition and rank, i.e.~writing fermionic
and bosonic pure states as sums of orthogonal Slater determinants or permanents, respectively.

For the case of the smallest non-trivial Hilbert space we were able to identify interesting properties
shared by all these systems: an appropriate
time reversal operator allows to define a {\it correlation measure}, the {\it concurrence}, which
can also be expressed through the coefficients of a state in the so-called {\it magic basis}. This
also allows for perfect analogues of Wootters' formula \cite{Woo:98} for the correlation measure of
mixed two-qubit states. As we noted, these analogies
follow directly if the system of two fermions in a four-dimensional single-particle space
is considered as the most general one from which the other two can be derived.

Subsequently we analyzed systems of fermions or bosons in higher-dimensional
single-particle spaces and systems of more than two indistinguishable
particles. There is no straight-forward extension of the results obtained for
low-dimensional Hilbert spaces. We gave necessary and sufficient
criteria to identify the Slater rank of pure two-particles states by contractions
of the $\varepsilon$-tensor with its coefficient matrix. These results are sufficient to classify
quantum correlations of pure states in arbitrary-dimensional single-particle spaces. For the case
of three or more indistinguishable particles it was first established that a general state cannot
be written as a sum of orthogonal Slater determinants, which demonstrates yet another analogy to
distinguishable particles. We showed that non-correlated pure states can be identified through
projections of the multi-particle state onto states of two fermions.
 
For general two-fermion mixed states we reviewed the concept of witnesses of Slater number $k$
and introduced a similar concept for bosons. In both cases we showed how to optimize
these witnesses and gave an example of an optimal witness.

Finally we explored the role of the PPT criterion
in bosonic systems. We gave separability criteria for low-rank two-boson mixed states in
a three-dimensional single-particle space having a positive partial transpose. We also showed
that there are no bosonic three-qubit states that have a positive partial transpose and nevertheless
contain quantum correlations. The same is true for bosonic $N$-qubit states of rank $\le N$. 

\section{Acknowledgments}
We wish to thank G.\ Birkl, J.I.\ Cirac, F.\ Hulpke, A.J.\ Leggett, D.\ Loss, M.\ Ku\'s, J.\ Mompart,
A.\ Sanpera, X.X.\ Yi and P.\ Zanardi for discussions. This work has been supported by
the Deutsche Forschungsgemeinschaft (SFB 407 and Schwerpunkt ``Quanteninformationsverarbeitung''and grant SCHL 539/1-1), EQUIP, the ESF PESC Programm on Quantum Information and the Welch Foundation.

\begin{appendix}

\section{Mode correlations}
\label{app:Zanardi}
Here we discuss briefly the rather complementary concept to deal with fermionic correlations
introduced by Zanardi \cite{Zan:01}.
Again consider the quantum dot example and suppose the system evolves to a state
(here we neglect interaction of the particles)
\begin{eqnarray}
\Ket{{\tilde \psi}'}&=&\frac1{2}\left[\,\Ket{\phi}\Ket{\up}\otimes
\Ket{\phi}\Ket{\down}+\Ket{\chi}\Ket{\up}\otimes\Ket{\chi}\Ket{\down}+
\Ket{\phi}\Ket{\down}\otimes
\Ket{\phi}\Ket{\up}+\Ket{\chi}\Ket{\down}\otimes\Ket{\chi}\Ket{\up}\,\right]\\
&=&\frac{1}{\sqrt2}\left[\fc{\phi,\up}\fc{\phi,\down}+\fc{\chi,\up}\fc{\chi,\down}\right]\Ket{0}
\end{eqnarray}
If the tunneling barrier is raised in this state then either Alice or Bob find to electrons
in their dot, i.e.\ there is no particle entanglement present here. But if we forget about the spin
and map the two particle basis state $\fc{i}\fc{j}\Ket{\0}$
onto $\Ket{n_{\phi}}\otimes\Ket{n_{\chi}}$, where $n_{\phi}$ and $n_{\chi}$ are
the occupation numbers of the left and the right dot, respectively, then the result is
a state $\propto\Ket{2}\otimes\Ket{0}+\Ket{0}\otimes\Ket{2}$. This is an
entangled state, but now the entangled entities are not the particles
themselves but "modes". Each mode is associated with (in this case more than one)
creation operator $\fc{i}$.

This is the concept of mode entanglement. In \cite{Zan:01,Zan:02}
correlations are analyzed not by looking at the 
particles themselves but by partitioning the state space into
different modes. In the simplest case these modes are associated with
the $2K$ single particle basis states such that the occupation number is
either $0$ or $1$ in the case of fermions.
These $2K$ modes are then mapped onto the space of $2K$ qubits such that
the $k^{{\rm th}}$ qubit is in state $\Ket{1}$ (state $\Ket{0}$) if the
$k^{{\rm th}}$ mode is occupied (not occupied).

More formally this mapping is describe for the basis states of $N$ fermions as
\cite{Zan:01}
\begin{eqnarray}
\Lambda:\qquad\hil_{2K}^F(N)&\;\rightarrow&\;(\C^2)^{\otimes 2K}\\
f_{i_1}^{\dagger}\ldots f_{i_N}^{\dagger}\Ket{\0}&\;\rightarrow&\;\otimes_{k=1}^{2K}
\Ket{\chi(k;\,i_1,\ldots,i_N)}
\end{eqnarray}
(for $i_1<\ldots<i_N$)
and linearly extended to the complete Hilbert space.
The characteristic function $\chi$ is defined as
\begin{equation}
\chi(k;\,i_1,\ldots,i_N)=\left\lbrace
\begin{array}{cc}
1\quad&k\in\set{i_1,\ldots,i_N}\\
0\quad&k\not\in\set{i_1,\ldots,i_N}
\end{array}
\right.
\end{equation}

The mapping is not bijective in this form as not all possible $K$-qubit
states are realized, i.e. $\hil_{2K}^F(N)$
is not isomorphic to this space. An isomorphism can be formed by
not fixing the number of fermions in the system, i.e. by considering
\begin{equation}
\Lambda:\hil_{2K}^F\rightarrow(\C^2)^{\otimes 2K}
\end{equation}
where $\hil_{2K}^F=\oplus_{N=1}^{2K} \hil_{2K}^F(N)$ is the total Fock space of
$0,\ldots,2K$ fermions. Notice that this point might raise problems
in concrete applications as it requires sources and drains to adjust the number
of particles.
%
%
%
%
\section{Proofs (Quantum correlations of pure fermionic and bosonic states in higher-dimensional Hilbert spaces)}
Here we give proofs of some of the lemma on indistinguishable 
particles in higher-dimensional Hilbert spaces from section \ref{Kai}.

\subsection{Fermionic states}
\label{proofs:fermionic}

First notice that according to equation (\ref{z}) for any complex
antisymmetric $2K\times2K$ coefficient matrix $w$ there is a
unitary transformation ${\cal U}$ such that
\begin{equation}
w'_{ij}=(UwU^T)_{ij}=
z_1\e^{ij34\ldots(2K)} + z_2\e^{12ij56\ldots(2K)} + \ldots +
z_{K}\e^{12\ldots ij}
\end{equation}
where $\e^{i_1,\ldots i_{2K}}$ is the totally antisymmetric unit tensor in
$2K\times\ldots\times2K$ and $z_k\geq0$.

\paragraph{Proof of lemma \ref{lem:2ferm6dim}:}
[$\Rightarrow$] Consider the case
$\alpha=5$, $\beta=6$. In the sum
all terms with at least one $i,j,k,l\in\left\{5,6\right\}$ vanish, and thus
\begin{equation}
\sum_{i,j,k,l}^6w_{ij}w_{kl}\e^{ijkl56}=
\sum_{i,j,k,l}^4w_{ij}w_{kl}\e^{ijkl}=C\left(\Ket{v}\right)
\end{equation}
where $\Ket{v}=v_{ij}f_i^{\dagger}f_j^{\dagger}\Ket{\0}\in{\cal A}({\cal H}_4\otimes{\cal H}_4)$ and $v_{ij}$
is constructed from $w_{ij}$ by deleting the last two columns and rows.

Because by assumption $\Ket{w}$ has Slater rank 1 also $\Ket{v}$ has to have
Slater rank 1 unless $\Ket{v}=0$. In both cases $C\left(\Ket{v}\right)$ has to vanish.
The same argument can be repeated for every combination of $\alpha$ and $\beta$
(only deleting different rows and columns from the coefficient matrix $w$).

\noindent [$\Leftarrow$] We will show the negated claim: there is at least one sum that does
not vanish if $\Ket{w}$ has Slater rank $>1$. The two possible cases to be
distinguished are Slater rank three or Slater rank two. If the Slater rank
is three then $w$ has full rank and thus $\det(w)\neq0$.
Using the expression for the determinant of an antisymmetric matrix from
equation (\ref{eqn:asdet}) it holds
\begin{equation}
48\det(w)=\sum_{\alpha,\beta=1}^6w_{\alpha\beta}
\underbrace{\sum_{i,j,k,l=1}^6w_{ij}w_{kl}\e^{ijkl\alpha\beta}}_{x^{\alpha\beta}}\neq 0,
\end{equation}
implying that at least one $x^{\alpha\beta}\neq 0$.

If $\Ket{w}$ has Slater rank 2 then $\rank{w}=4$ and we can find a unitary transformation
${\cal U}$ such that
\begin{equation}
w'_{i'j'}=\sum_{i,j=1}^6 U_{i'i}w_{ij}U_{j'j}
\end{equation}
contains only zeros in the last two rows and columns. $\Ket{w}$ has Slater rank 2
if and only if $\Ket{w'}=\sum_{i,j} w'_{ij}(f')_i^{\dagger}(f')_j^{\dagger}\Ket{\0}$ has Slater rank 2
and it follows ($w'$ is a $4\times4$ matrix only and thus it is possible to use the 
concurrence from section \ref{analogs} which does not vanish if and only if $\Ket{w'}$ has Slater rank 2)
\begin{eqnarray}
0\neq C\left(\Ket{w'}\right) &=& \sum_{i,j,m,n=1}^4w'_{ij}w'_{mn}\,\e^{ijmn}\\
&=& \sum_{i,j,m,n=1}^6w'_{ij}w'_{mn}\,\e^{ijmn56}\\
&=& \mathop{\sum_{i,j,m,n=1}^6}_{i',j',m',n'=1} U_{ii'}\, w_{i'j'}\, U_{jj'}\,
U_{mm'} w_{m'n'} U_{nn'}\,\e^{ijmn56}\\
&=& \mathop{\sum_{i,j,m,n=1}^6}_{i',j',m',n'=1} w_{i'j'}\, w_{m'n'}\,
\underbrace{U_{ii'}\, U_{jj'}\, U_{mm'}\, U_{nn'}\,\e^{ijmn56}}_{\widetilde\e^{\,i'j'm'n'}}
\end{eqnarray}
It is easy to verify that $\widetilde\e^{\,i'j'm'n'}$ is also antisymmetric
in all indices so it has to have the form
\begin{equation}
\widetilde\e^{\,i'j'm'n'}=
\xi_{56}\e^{i'j'm'n'56}+\xi_{46}\e^{i'j'm'n'46}+\ldots+\xi_{12}\e^{i'j'm'n'12}
\end{equation}
(where some but not all $\xi_{\alpha\beta}$ may be zero) and then
\begin{eqnarray}
0 & \neq & \sum_{i',j',m',n'=1}^6 w_{i'j'} w_{m'n'}
\left(\xi_{56}\e^{i'j'm'n'56}+\xi_{46}\e^{i'j'm'n'46}+\ldots+\xi_{12}\e^{i'j'm'n'12}\right)\\
0 & \neq & \sum_{\alpha\beta=1}^6 \xi_{\alpha\beta} \underbrace{\sum_{i',j',k',l'=1}^6\e^{i'j'k'l'\alpha\beta}w_{i'j'}w_{k'l'}}_{x^{\alpha\beta}}
\end{eqnarray}
such that again at least one $x^{\alpha\beta}\neq0$.

%
%
%

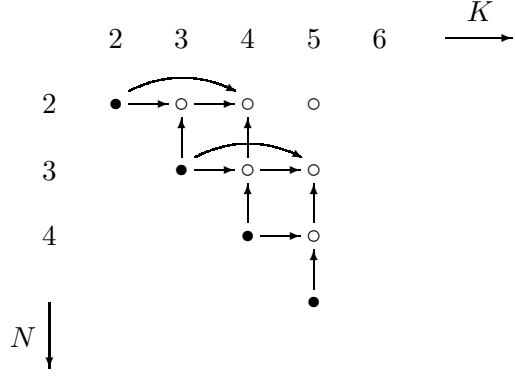
\begin{figure}[t]
\begin{center}
\begin{picture}(200,150)(0,-150)
\put(25,-50){\makebox(0,0){$2$}}
\put(25,-75){\makebox(0,0){$3$}}
\put(25,-100){\makebox(0,0){$4$}}
\put(25,-125){\vector(0,-1){25}}
\put(15,-138){\makebox(0,0){$N$}}

\put(50,-25){\makebox(0,0){$2$}}
\put(75,-25){\makebox(0,0){$3$}}
\put(100,-25){\makebox(0,0){$4$}}
\put(125,-25){\makebox(0,0){$5$}}
\put(150,-25){\makebox(0,0){$6$}}
\put(175,-25){\vector(1,0){25}}
\put(188,-15){\makebox(0,0){$K$}}

\put(50,-50){\circle*{4}}
\put(75,-75){\circle*{4}}
\put(100,-100){\circle*{4}}
\put(125,-125){\circle*{4}}
\put(75,-50){\circle{4}}
\put(100,-50){\circle{4}}
\put(100,-75){\circle{4}}
\put(125,-50){\circle{4}}
\put(125,-75){\circle{4}}
\put(125,-100){\circle{4}}

\put(75,-70){\vector(0,1){15}}
\put(55,-50){\vector(1,0){15}}

\put(100,-95){\vector(0,1){15}}
\put(80,-75){\vector(1,0){15}}

\put(100,-70){\vector(0,1){15}}
\put(80,-50){\vector(1,0){15}}
\qbezier(55,-45)(75,-35)(95,-45)
\put(93,-44){\vector(2,-1){3}}

\put(125,-120){\vector(0,1){15}}
\put(105,-100){\vector(1,0){15}}

\put(125,-95){\vector(0,1){15}}
\put(105,-75){\vector(1,0){15}}
\qbezier(80,-70)(100,-60)(120,-70)
\put(118,-69){\vector(2,-1){3}}
\end{picture}
\caption{Construction of the proof of lemma \ref{lem:2FmSrN} by induction. Black dots mark the combinations
[K=N,N] where equation (\ref{eqn:asdet}) can be applied.} 
\label{fig:constructProof}
\end{center}
\end{figure}

\paragraph{Proof of lemma \ref{lem:2FmSrN}:}
If $N=K$ then the claim is that $\Ket{w}$ does not have maximal Slater rank
if and only if
\begin{equation}
\sum_{i_1,\ldots,i_{2K}=1}^{2K}w_{i_1i_2}\ldots w_{i_{2K-1}i_{2K}}
\e^{i_1\ldots i_{2K}}=0
\end{equation}
This is the statement
of equation (\ref{eqn:asdet}) so lemma \ref{lem:2FmSrN} surely holds for all cases
$[K=N,N]$. This is the starting point for giving the proof by induction. For a
combination $[K,N]$ ($N<K$) the assumption will be that the lemma holds for
$[K,N+1]$, $[K-1,N]$ and $[N,N]$ (figure \ref{fig:constructProof} illustrates
the induction process).

\noindent[$\Rightarrow$] Introduce a two-fermion state
$\Ket{v}\in \as{\hil_{2K-2}\otimes\hil_{2K-2}}$ by constructing its 
coefficient matrix $v$ from $w$ by deleting the rows and columns
$\alpha_{2(K-N)}$ and $\alpha_{2(K-N)-1}$.
Since $\Ket{w}$ is assumed to have Slater rank $<N$ also $\Ket{v}$
has to have Slater rank $<N$ (because its mathematical rank cannot raise
when some rows and columns are deleted). Then it follows
\begin{eqnarray}
\sum_{i_1,\ldots,i_{2N}=1}^{2K}w_{i_1i_2}\ldots w_{i_{2N-1}i_{2N}}
\e^{i_1\ldots i_{2N}\,\alpha_1\ldots\alpha_{2(K-N)}}&=&\\
\qquad\sum_{i_1,\ldots,i_{2N}=1}^{2(K-1)}v_{i_1i_2}\ldots v_{i_{2N-1}i_{2N}}
\e^{i_1\ldots i_{2N}\,\alpha_1\ldots\alpha_{2(K-1-N)}}&=&0,
\end{eqnarray}
because by assumption the lemma holds for $K-1$ and $N$.

\noindent [$\Leftarrow$] We will show that there is at least one non-vanishing sum
if the Slater rank is $\geq N$. Then if all terms vanish the Slater rank has
to be $<N$. Again there are two cases to be distinguished:

{\bf Case 1}: $\Ket{w}$ has Slater rank $>N$. By assumption the lemma holds for
$[K,N+1]$ and thus
\begin{equation}
0\neq\sum_{i_1,\ldots,i_{2N+2}=1}^{2K}w_{i_1i_2}
\ldots w_{i_{2N+1}i_{2N+2}}\e^{i_1\ldots i_{2N+2}\alpha_1
\ldots\alpha_{2(K-N-1)}}
\end{equation}
for at least one combination of $\alpha_1,\ldots,\alpha_{2(K-N-1)}$. But then
\begin{equation}
0\neq\sum_{i_{2N+1},i_{2N+2}}^{2K}w_{i_{2N+1}i_{2N+2}}
\underbrace{\sum_{i_1,\ldots,i_{2N}}^{2K}
w_{i_1i_2}\ldots w_{i_{2N-1}i_{2N}}\e^{i_1\ldots i_{2N}i_{2N+1}i_{2N+2}\alpha_1
\ldots\alpha_{2(K-N-1)}}}_{x^{i_{2N+1}i_{2N+2}\alpha_1\ldots\alpha_{2(K-N-1)}}}
\end{equation}
such that at least one $x^{i_{2N+1}i_{2N+2}\alpha_1\ldots\alpha_{2(K-N-1)}}\neq0$.

{\bf Case 2}: $\Ket{w}$ has Slater rank $N$. Then $\rank{w}=2N$ and there is a
unitary transformation $U\in SU(2K)$ such that
\begin{equation}
w'_{i'j'}=\sum_{i,j=1}^{2K}U_{i'i}w_{ij}U_{j'j}
\end{equation}
contains zeros in the last $(2K-2N)$ columns and rows. $\Ket{w'}$ also has to have
Slater rank N and because the lemma is assumed to be valid for $[K=N,N]$ 
it holds
\begin{eqnarray}
0&\neq& \sum_{i_1,\ldots,i_{2N}=1}^{2N}w'_{i_1i_2}\ldots w'_{i_{2N-1}i_{2N}}
\e^{i_1\ldots i_{2N}}\\
&=& \sum_{i_1,\ldots,i_{2N}=1}^{2K}w'_{i_1i_2}\ldots w'_{i_{2N-1}i_{2N}}
\e^{i_1\ldots i_{2N}(2N+1)\ldots(2K)}\\
&=& \mathop{\sum_{i_1\ldots i_{2N}=1}^{2K}}_{j'_1\ldots j'_{2N}=1}
U_{i_1j'_1}w_{j'_1j'_2}U_{i_2j'_2}\ldots
U_{i_{2N-1}j'_{2N-1}}w_{j'_{2N-1}j'_{2N}}U_{i_{2N}j'_{2N}}
\e^{i_1\ldots i_{2N}(2N+1)\ldots(2K)}\\
&=& \sum_{j'_1\ldots j'_{2N}=1}^{2K} w_{j'_1j'_2} \ldots w_{j'_{2N-1}j'_{2N}} \underbrace{
\sum_{i_1\ldots i_{2N}=1}^{2K}U_{i_1j'_1} \ldots U_{i_{2N}j'_{2N}}
\e^{i_1\ldots i_{2N}(2N+1)\ldots(2K)}}_{
\widetilde\e^{\,j'_1\ldots j'_{2N}(2N+1)\ldots(2K)}}.
\end{eqnarray}
$\widetilde\e^{\,j'_1\ldots j'_{2N}(2N+1)\ldots(2K)}$ has to be antisymmetric in all indices
so it has the form
\begin{equation}
\widetilde\e^{\,j'_1\ldots j'_{2N}(2N+1)\ldots(2K)}=
\sum_{\alpha_1,\ldots,\alpha_{2(K-N)}=1}^{2K}\xi_{\alpha_1\ldots\alpha_{2(K-N)}}
\e^{j'_1\ldots j'_{2N}\alpha_1\ldots\alpha_{2(K-N)}}
\end{equation}
(some but not all $\xi$ might be zero) and then
\begin{equation}
0\neq\sum_{\alpha_1\ldots\alpha_{2(K-N)}=1}^{2K} \xi_{\alpha_1\ldots\alpha_{2(K-N)}} \underbrace{
\sum_{j'_1\ldots j'_{2N}=1}^{2K} w_{j'_1j'_2} \ldots w_{j'_{2N-1}j'_{2N}}
\e^{j'_1\ldots j'_{2N}\alpha_1\ldots\alpha_{2(K-N)}}
}_{x^{\alpha_1\ldots\alpha_{2(K-N)}}},
\end{equation}
where at least one $x^{\alpha_1\ldots\alpha_{2(K-N)}}\neq0$ is necessary.

%
%

\paragraph{Proof of lemma \ref{lem:3FmSr1}:} 
[$\Rightarrow$] $\Ket{w}$ has Slater rank one so there exists $U$,
$UU^{\dagger}=1$, such that
\begin{equation}\label{eqn:wtransform}
w'_{ijk}=w_{lmn}\;U_{li}\;U_{mj}\;U_{nk}=z\e^{ijk456\ldots 2K}.
\end{equation}
Then, for arbitrary $\Ket{a}$ and writing $\Ket{a'}=U^{\dagger}\Ket{a}$, we have
\begin{equation}\label{eqn:wtov}
\Ket{v'}=R_{\Ket{a'}}\Ket{w'}\quad\mbox{with}\;
v'_{ij}=3 \sum_{k=1}^{2K}w'_{ijk}a'_k=
\sum_{k,l=1}^{2K}z\e^{ijk456\ldots 2K}a_lU^{*}_{kl}=
(A\oplus0)_{ij}
\end{equation}
with the $3\times3$ antisymmetric matrix $A$ and $0$ as the
$(2K-3)\times (2K-3)$-dimensional
zero. $A$ has rank zero or two and thus there exists
unitary $V$ such that
\begin{equation}
(Vv'V^T)_{ij}=\beta\e^{ij345\ldots(2K)}
\end{equation} 
for some $\beta$. Inserting equations (\ref{eqn:wtransform}) and (\ref{eqn:wtov}) shows that
$\Ket{v}=R_{\Ket{a}}\Ket{w}$ indeed has Slater rank one (if $\beta\neq0$):
\begin{eqnarray}
\beta\e^{ij345\ldots(2K)} & = & 
\sum_{l,m,n,o,p,q=1}^{2K}V_{im}\,w_{opq}\,U_{mo}U_{np}U_{kq}\,
a_l\,U^{*}_{kl}\,V_{jn}\\
& = & \sum_{o,p,q=1}^{2K}(U^{T}V)_{oi}\,w_{opq}\,a_q\,(U^{T}V)_{pj}.
\end{eqnarray}

\noindent [$\Leftarrow$]
Here we make use of the fact that an $N$-fermion state $\Ket{w}$ (coefficient matrix $w$) in a 
$2K$-dimensional single-particle space 
has Slater rank one (it can, after some transformation of the single-particle
basis, be written as a single Slater determinant) if and only if there
exists a set of orthonormal vectors $\set{\vc{e}^{\alpha}}_{\alpha=1,\ldots,K}$
with $\vc{e}^{\alpha}\cdot\vc{e}^{\beta}=\delta^{\alpha\beta}$, such that
for $\alpha_1,\ldots,\alpha_N\in\set{1,\ldots,2K}$
\begin{equation}\label{eqn:slateRankOneBasis}
w_{ijk}\,e^{\alpha_1}_i\,\ldots\,e^{\alpha_N}_k=ze^{\alpha_1\ldots\alpha_N(N+1)\ldots(2K)}.
\end{equation}
These vectors are the columns of the matrix $U$ that transforms the single
particle space.

Choose an arbitrary $\vc{e}^{\,1}$, $|\vc{e}^{\,1}|=1$, such that 
$v_{jk}=\sum_i w_{ijk} e^{\,1}_i$ is not the $2K\times2K$-dimensional zero. 
The resulting two fermion state $\Ket{v}$ by assumption has Slater rank 1
and thus there exist $\vc{e}^{\,2},\vc{e}^{\,3},\ldots,\vc{e}^{\,2K}$ such that
$\vc{e}^{\,\alpha}\vec{e}^{\,\beta}=\delta^{\alpha\beta}$ and
\begin{equation}
\sum_{j,k=1}^{2K}v_{jk}e^{\,\alpha}_je^{\,\beta}_k=
\sum_{i,j,k=1}^{2K}w_{ijk}e^1_ie^{\,\beta}_je^{\,\gamma}_k
= z\e^{1\beta\gamma4\ldots(2K)}.\label{eqn:3fm1}
\end{equation}
But we furthermore claim that $\Ket{w}$ has Slater rank one:
\begin{equation}
\sum_{i,j,k=1}^{2K}w_{ijk}e^{\,\alpha}_ie^{\,\beta}_je^{\,\gamma}_k
= z\e^{\alpha\beta\gamma4\ldots(2K)}
\quad\mbox{for}\quad\alpha,\ \beta,\ \gamma\in\left\{1\ldots2K\right\}.
\end{equation}
Because of equation (\ref{eqn:3fm1}) this is obvious 
if $(\alpha,\,\beta,\,\gamma)$ is a permutation of $(1,\,2,\,3)$ and otherwise if $\alpha=1$.
So further attention has only to be paid to the cases where $1<\alpha<\beta<\gamma$.

 Consider 
$\Ket{v^{\alpha}}$ with $v^{\alpha}_{jk}=\sum w_{ijk}e_k^{\,\alpha}$ for $1<\alpha\leq3$.
By assumption the two-fermion state $\Ket{v^{\alpha}}$ has Slater rank one (or is zero)
and thus there is at most a two-dimensional subspace ${\cal M}$ such that for any 
two $\vc{a}\in{\cal M}$ and $\vc{b}\in{\cal M}$, $\vc{a}\neq\vc{b}$, 
$\sum w_{ijk}e_i^{\,\alpha}a_jb_k\neq0$. Because of eqn. (\ref{eqn:3fm1}) ${\cal M}$ is
spanned by $\vc{e}^{\,\beta},\ \vc{e}^{\,\gamma}$  where $\left\{\beta,\ \gamma\right\}=
\left\{1,2,3\right\}\setminus\left\{\alpha\right\}$. The $\vc{e}\,$'s are orthogonal and thus
$w_{ijk}e^{\,\alpha}_ie^{\beta'}_je^{\gamma'}_k=0$ if $1<\alpha<4$ and $\alpha<\beta'<\gamma'$.

For $3<\alpha<\beta<\gamma$ fixed suppose 
$\sum w_{ijk}e^{\alpha}_ie^{\beta}_je^{\gamma}_k\neq 0$.
Then there exist a two-dimensional subspace ${\cal M}$ spanned by $\vc{e}^{\,2}$ and
$\vc{e}^{\,3}$ and an orthogonal subspace ${\cal N}$ spanned by $\vc{e}^{\,\beta}$
and $\vc{e}^{\,\gamma}$ such that
$\sum w_{ijk}\left[e_i^{\,1}+e_i^{\,\alpha}\right]a_jb_k\neq0$ for 
$\vc{a}$ and $\vc{b}$ both in ${\cal M}$ \emph{or} both in ${\cal N}$ and 
$\vc{a}\neq\vc{b}$. This contradicts the assumption that $R_{\Ket{e^1}+\Ket{e^{\alpha}}}\Ket{w}$
has Slater rank one.
%
%

\paragraph{Proof of lemma \ref{lem:NFmSr1}:}
[$\Rightarrow$] $\Ket{w}$ has Slater rank one, i.e. we can find
$\vc{e}^{\,1},\ldots,\vc{e}^{\,2K}$ such that
\begin{equation}\label{propsl1}
\sum_{i_1,\ldots,i_N=1}^{2K} w_{i_1\ldots i_N}e_{i_1}^{\,\alpha_1}
\ldots e_{i_N}^{\,\alpha_N}
= z \e^{\alpha_1\ldots\alpha_N\; (N+1)\ldots (2K)}
\end{equation}
Now consider $v_{i_1\ldots i_{N-1}}=\sum w_{i_1\ldots i_N}x_{i_N}$for arbitrary
$\vc{x}$. If $\vc{x}=\vc{e}^{\,\alpha}$ then $\Ket{v}$ has Slater rank 1
if $\alpha\leq N$ or is zero if $\alpha>N$. Otherwise it holds that
\begin{equation}
\sum_{i_{N-1}=1}^{2K}v_{i_1\ldots i_{N-1}}e^{\,\alpha}_{i_{N-1}}=0\quad
\mbox{for}\;\alpha>N\qquad\mbox{and}\qquad
\sum_{i_{N-1}=1}^{2K}v_{i_1\ldots i_{N-1}}x_{i_{N-1}}=0
\end{equation}
and we can choose $\vc{b}^{\,\alpha}=\vc{e}^{\,\alpha}$ for $\alpha>N$,
$\vc{b}^N=\vc{x}$ and the remaining $\vc{b}^{\,\alpha}$ for $\alpha<N$ such that
$\vc{b}^{\,\alpha_1}\vc{b}^{\,\alpha_2}=\delta^{\alpha_1\alpha_2}$ to get
\begin{equation}
\sum_{i_1,\ldots,i_{N-1}=1}^{2K} v_{i_1\ldots i_{N-1}}b_{i_1}^{\,\alpha_1}
\ldots b_{i_{N-1}}^{\,\alpha_{N-1}}
= \widetilde z\, \e^{\alpha_1\ldots\alpha_{N-1}\, N\ldots (2K)}
\end{equation}
which means that $\Ket{v}$ has Slater rank one if $\widetilde z\neq 0$ and 
is zero otherwise.

\noindent [$\Leftarrow$]: similar to the three-fermion case.

\subsection{Bosonic states}\label{proofs:bosonic}

\paragraph{Proof of lemma \ref{lem:2bos3dim}:}
[$\Rightarrow$] Let $\Ket{v}$ have Slater rank one and for any $1\leq\alpha\leq3$ let $v'$ be
the $2\times2$-matrix that results from deleting the row and column $\alpha$.
If $\Ket{v}$ has Slater rank one then also $\Ket{v'}=
\sum_{i,j=1}^2 v'_{ij}\bc{i}\bc{j}\Ket{\0}$ has Slater rank one (unless it is zero)
and the claim follows from equation (\ref{eqn:conbos}).

\noindent [$\Leftarrow$] We will show that at least one sum in equation (\ref{eqn:2bos3dim})
does not vanish if $\Ket{v}$ has Slater rank larger than one. The two possible cases are
Slater rank three and Slater rank two.
If the Slater Rank is three then $v$ has full rank and $\det v\neq 0$. There exists
a unitary transformation $U$ of the single particle space such that
\begin{equation}
v'_{i'j'}=\sum_{i,j=1}^3U_{ii'}v_{i'j'}U_{jj'}
\end{equation}
is diagonal. In this basis
\begin{eqnarray}
0\neq 3!\,\det v' & = & \sum_{i=1}^{3}v'_{ii}\sum_{k,m=1}^{3}v'_{kk}v'_{mm}\e^{ikm}\,\e^{ikm}\\
& = & \sum_{i=1}^3 v'_{ii}\mathop{\sum_{k,m=1}^3}_{k',l',m',n'=1}
U_{kk'}v_{k'l'}U_{kl'}U_{mm'}v_{m'n'}U_{mn'}\e^{ikm}\,\e^{ikm}\\
& = &\sum_{i=1}^3 v'_{ii}\sum_{\alpha,\beta=1}^3\xi_{\alpha}^i\xi_{\beta}^i
\underbrace{\sum_{k',l',m',n'=1}^3
v_{k'l'}v_{m'n'}\e^{\alpha k'm'}\e^{\beta l'n'}}_{x^{\alpha\beta}}\neq 0
\end{eqnarray}
and thus at least one $x^{\alpha\beta}\neq0$. To show that his implies that at least one 
$x^{\alpha\alpha}\neq0$, let $\set{\alpha,\beta}\in\set{1,2,3}$ and let $\gamma$
be the remaining index. Then
\begin{eqnarray}
|x^{\alpha\beta}|&=&2 |\e^{\alpha\beta\gamma}\,\e^{\beta\alpha\gamma}
v_{\beta\alpha}v_{\gamma\gamma}+
e^{\alpha\beta\gamma}\,\e^{\beta\gamma\alpha}v_{\beta\gamma}v_{\gamma\alpha}|\\
&=&2|\e^{\alpha\beta\gamma}\,\e^{\beta\alpha\gamma}v_{\beta\alpha}v_{\gamma\gamma}-
e^{\alpha\beta\gamma}\,\e^{\beta\alpha\gamma}v_{\beta\gamma}v_{\gamma\alpha}|\\
&=&2|v_{\beta\alpha}v_{\gamma\gamma}-v_{\beta\gamma}v_{\gamma\alpha}|\must0
\quad\Leftrightarrow\quad
1=v_{\gamma\gamma}^2\frac{v_{\beta\alpha}^2}{v_{\gamma\alpha}^2 v_{\beta\gamma}^2}.
\label{eqn:condxabzero}
\end{eqnarray}
On the other hand if $\alpha\in{1,2,3}$ and $\beta,\gamma$ are the remaining two indices,
then 
\begin{eqnarray}
|x^{\alpha\alpha}|&=&2|\e^{\alpha\beta\gamma}\,\e^{\alpha\beta\gamma}v_{\beta\beta}v_{\gamma\gamma}+
\e^{\alpha\beta\gamma}\,\e^{\alpha\gamma\beta}v_{\beta\gamma}^2|\\
&=&|2v_{\beta\beta}v_{\gamma\gamma}-v_{\beta\gamma}^2|\must0\quad\Leftrightarrow\quad
 v_{\beta\gamma}^2=v_{\beta\beta}v_{\gamma\gamma}
\end{eqnarray}
and thus if $x^{11}=0$, $x^{22}=0$ and $x^{33}=0$ then $v_{23}^2=v_{22}v_{33}$,
$v_{13}^2=v_{11}v_{33}$ and $v_{12}^2=v_{11}v_{22}$. Then equation (\ref{eqn:condxabzero})
is fulfilled and $x^{\alpha\beta}=0$ for all choices of $\alpha$ and $\beta$. Thus at least one
$x^{\alpha\beta}\neq0$ implies that at least one $x^{\alpha\alpha}\neq0$.

If the Slater Rank is two then there exists a unitary transformation $U$ such that
\begin{equation}
v'_{i'j'} = \sum_{i,j=1}^3U_{ii'}v_{i'j'}U_{jj'}=\rm{diag}(v'_{11},v'_{22},0),
\end{equation}
and thus
\begin{eqnarray}
0&\neq&\sum_{k,m=1}^2 v'_{kk}v'_{mm}\e^{3km}\,\e^{3km}\\
&=&\sum_{\substack{k,m=1\\k',l',m',n'=1}}^3U_{kk'}v_{k'l'}U_{kl'}U_{mm'}v_{m'n'}U_{mn'}\e^{3km}\,\e^{3km},
\end{eqnarray}
such that the above arguments can be repeated. Using this, lemma \ref{lem:2bosKdim} can then be proven
by induction.

\paragraph{Proof of lemma \ref{lem:NbosSr1}:}
We will give the proof for the case $N=3$. [$\Rightarrow$]: $\Ket{v}$ has Slater rank one and,
following definition \ref{def:nonCorrNBos}, there exists a unitary transformation $U$ such that
\begin{equation}
v'_{ijk}=\sum_{l,m,n=1}^K
v_{lmn}U_{il}U_{jm}U_{kn}=z\,\delta_{i1}\delta_{j1}\delta_{k1}.
\end{equation}
Then for arbitrary $\Ket{a}$ it is obvious that $R_{\Ket{a}}\Ket{v}$
has Slater rank one (unless it is zero), because (here $a'_k=U_{kl}a_l$)
\begin{equation}
3\sum_{k=1}^Kv'_{ijk}a'_k=a_1\,z\,\delta_{i1}\delta_{j1}.
\end{equation}

\noindent [$\Leftarrow$] Notice that if a $K\times K$-matrix 
$v$ has rank one and (normalized) eigenvector $\vc{e}$, then if the normalized 
vector $\vc{x}$ maximizes $\parallel v\,\vc{x}\parallel$ then $\vc{x}=\vc{e}$. 
Now let $\vc{x}$ and $\vc{y}$ be normalized vectors varying over the unit sphere 
and maximize $\parallel \sum_{ijk}v_{ijk}x_iy_j\parallel$. A maximum exists because 
the product of two unit spheres is a compact space and $\parallel . \parallel$ is a
continuous function. Let the maximum occur at
$\vc{x}=\vc{e}^{\,(1)}$ and $\vc{y}$. Now $\vc{y}$ is the eigenvector of
$\sum_i v_{ijk}e_i^{(1)}$, which by assumption has rank one, because $\vc{y}$ maximizes 
$\parallel \sum_{i,j}v_{ijk}e^{(1)}_iy_j\parallel$. Similarly $\vc{e}^{\,(1)}$ is the eigenvector of
$\sum_i v_{ijk}y_i$. Then we have
\begin{equation}
\sum_{i,j}^3v_{ijk} e^{(1)}_iy_j=\lambda_1y_k\qquad\text{and}\qquad
\sum_{i,j}^3v_{ijk}y_ie_j^{(1)}=\widetilde\lambda_1 e^{(1)}_k
\end{equation}
and by comparing both equations
\begin{equation}
\quad\vc{y}\equiv\vc{e}^{\,(1)},\quad\widetilde\lambda_1\equiv\lambda_1,
\quad\sum_{i,j,k}^3v_{ijk}e_i^{(1)}e_j^{(1)}e_k^{(1)}=\lambda_1,
\end{equation}
where $\lambda_1$ can be chosen real and positive by adjusting the phase of $\vc{e}^{\,(1)}$.
Furthermore $\sum_{i,j}v_{ijk}e_i^{(1)}z_j=0$
for any $\vc{z}\perp\vc{e}^{(1)}$.

Now repeat this maximization, but this time restrict $\vc{x}$ and $\vc{y}$ 
to the space perpendicular to $\vc{e}^{(1)}$. If the result
is called $\vc{e}^{(2)}$, then $\sum_{i,j,k}v_{ijk}e_i^{(2)}e_j^{(2)}e_k^{(2)}=\lambda_2$ and
$\lambda_2\leq \lambda_1$. In this way a complete basis is constructed.
If $v$ is written in this basis then
\begin{equation}
v'=v\frac{1}{\sqrt{K}}\left(\vc{e}^{\,(1)}+\ldots+\vc{e}^{\,(K)}\right)=
\frac{1}{\sqrt{K}}\text{diag}\left[\lambda_1,\ldots,\lambda_K\right],\quad
\lambda_1\geq\lambda_2\geq\ldots\geq\lambda_K\geq0
\end{equation}
By assumption $\Ket{v'}$ has to have Slater rank one, i.e.\ 
$v'$ has to have rank one, and it follows that $\lambda_2=\ldots=\lambda_K=0$.
Thus 
\begin{equation}
\sum_{i,j,k}v_{ijk}e_i^{\,(\alpha)}e_j^{\,(\beta)}e_k^{\,{(\gamma)}}=\lambda_1
\delta_{1\alpha}\delta_{1\beta}\delta_{1\gamma}
\end{equation}
and $\Ket{v}$ has indeed Slater Rank 1.

\end{appendix}


\begin{thebibliography}{50}
\bibitem{note1} In Refs.~\cite{SCK+:00,SLM:00} the concurrence of
a two-fermion state was denoted by $\eta$. To stress the analogy between
the different kinds of systems sudied here we shall use the symbol
${\cal C}$ throughout this work.
\end{thebibliography}
\end{document}